\documentclass[aps,prd,showpacs,twocolumn]{revtex4}

\usepackage{epsfig}
\usepackage{graphicx}

\begin{document}

\title{Constraining the Little Higgs model of Schmaltz, Stolarski, and Thaler with recent results from the LHC} 

\author{Pat Kalyniak\footnote{kalyniak@physics.carleton.ca}}
\affiliation{Ottawa-Carleton Institute for Physics, Department of Physics, Carleton University, Ottawa, Canada K1S 5B6}

\author{Travis A. W. Martin\footnote{tmartin@triumf.ca}}
\affiliation{TRIUMF, 4004 Wesbrook Mall, Vancouver, Canada V6T 2A3}

\author{Kenneth Moats\footnote{kmoats@physics.carleton.ca}}
\affiliation{Ottawa-Carleton Institute for Physics, Department of Physics, Carleton University, Ottawa, Canada K1S 5B6}

\date{\today}

\begin{abstract}
In this paper, we use the latest Higgs measurements from ATLAS and CMS to constrain the parameter space of the model of Schmaltz, Stolarski and Thaler, a Little Higgs model with two Higgs doublets, which we will refer to as the BLH model. 
We account for all production and decay modes explored at ATLAS and CMS in two scenarios: a general case, which assumes the $h_0$ state is light ($m_{h_0} \approx 125$ GeV) and the masses of the other neutral scalars ($H_0$ and $A_0$) are allowed to vary, and a case with a near-degeneracy between the masses of the $h_0$ and $A_0$ and, for some choices of parameters, the $H_0$ states. The near-degeneracy scenario can result in an enhanced diphoton rate, as measured by ATLAS, but is largely ruled out by a combination of the $h_0 \rightarrow \tau^+\tau^-$ and the heavy $H_0 \rightarrow W^+W^-$ measurements. In the general case, we find large regions of parameter space that are in better agreement with either the ATLAS or CMS results than is the SM. However, a significantly enhanced diphoton rate is only possible through large contributions to the $h_0 \gamma \gamma$ effective coupling from charged Higgs bosons in a region of parameter space that borders on violation of perturbativity in the scalar sector.
\end{abstract}

\maketitle

\section{Introduction}

After analyzing the results from approximately 5~fb$^{-1}$ of integrated luminosity at both 7 TeV and 8 TeV centre-of-mass collision energies, both ATLAS and CMS revealed the discovery of a new resonance in the $\gamma\gamma$, $ZZ^*\rightarrow 4l$ and $WW^* \rightarrow l^+l^- \ensuremath{{\not\mathrel{E}}_T}$ decay channels, consistent with a Higgs boson at a mass of approximately 125 GeV, with a combined significance of more than $5\sigma$ \cite{Aad:2012tfa,Chatrchyan:2012ufa}. Now that the remaining 8 TeV collision data, corresponding to approximately 20~fb$^{-1}$ of integrated luminosity, has been analyzed, this new state continues to be consistent with a Higgs boson.  However, there are indications that its branching ratios might deviate from those of the Standard Model (SM) Higgs boson, particularly in the diphoton decay rate that is sensitive to the presence of new physics \cite{ATLAS:2012klq,CMS:yva}. More data will be needed for precise determination of the branching ratios.

Prior to Moriond 2013, both the CMS and the ATLAS experiments found an enhancement in the diphoton signal strength, without significant deviations from SM values in the $ZZ^*$ and $WW^*$ signal strengths \cite{Cheung:2013kla,ATLAS:2012znl,ATLAS:2012klq,ATLAS:2012dsy,ATLAS:2012bmv,CMS:aya,CMS:yxa,CMS:lng}. Following Moriond 2013, CMS updated their diphoton analysis with results that were in better agreement with the SM predictions \cite{CMS:ril}. Of interest, however, is that ATLAS \cite{ATLAS:2013sla,ATLAS:2013oma} still observes an excess in the diphoton rate at a significance of approximately $2\sigma$; and the ATLAS diphoton resonance provides a best fit invariant mass for the Higgs boson that is larger than the measured resonance mass for the $ZZ^*$ final state (126.8 GeV versus 124.3 GeV).

Beyond the Standard Model physics may be significantly constrained by comparing its predictions to the measured mass and the various measured signal strengths ($\hat{\mu}$) of the Higgs-like state. In general, vector boson fusion (VBF), vector boson associated production (VH), and the $ZZ^*$/$WW^*$ decay modes are sensitive to modifications of the Higgs boson couplings to gauge bosons; top quark associated production (ttH) and the fermion decay modes are sensitive to modifications of the Higgs boson couplings to fermions; and gluon fusion (ggF) and the diphoton ($\gamma\gamma$) and $Z\gamma$ loop-induced decay modes are sensitive to the presence of new coloured and electrically charged states, respectively, that couple to the Higgs boson, as well as modifications to the $hW^+W^-$ and $ht\bar{t}$ couplings.

In general, Little Higgs models without $T$-parity are more highly constrained by precision electroweak measurements than by the LHC Higgs results \cite{Reuter:2012sd,Han:2013ic}, while $T$-parity models are primarily constrained by relic abundance considerations \cite{Wang:2013yba} and LHC search results \cite{Reuter:2013iya}.  
The recent non-$T$-parity Little Higgs model of Schmaltz, Stolarski and Thaler, which we will refer to as the BLH model \cite{Schmaltz:2010ac},
is not as constrained by precision measurements due to the presence of a custodial symmetry and a disassociation of the masses of the top partner and heavy gauge boson states.  

The BLH model features a global $SO(6)_A\times SO(6)_B$ symmetry that is broken to a diagonal $SO(6)_V$ at a scale $f\sim\mathcal{O}$(TeV)  when a non-linear sigma field, $\Sigma$, develops a vev ($\langle\Sigma\rangle = 1\!\!1$). The resulting 15 pseudo-Nambu Goldstone bosons are parameterized as two real $SU(2)_L$ triplets, $\phi^a$ and $\eta^a$ ($a=1, 2, 3$), two complex Higgs doublets, $h_1$ and $h_2$, and a real singlet $\sigma$. A general two Higgs doublet potential is generated in part explicitly and in part radiatively, where the quartic coupling for the Higgs arises when integrating out the heavy scalar singlet, $\sigma$. A second global symmetry of the form $SU(2)_C\times SU(2)_D$ is also present, and is broken to a diagonal $SU(2)$ at a scale $F>f$ when a second non-linear sigma field, $\Delta$, develops a vev ($\langle\Delta\rangle = 1\!\!1$). To connect these two non-linear sigma models, the $SU(2)_{LA} \subset SO(6)_A$ and $SU(2)_C$ symmetries are gauged with the same $SU(2)_A$ gauge bosons, while the $SU(2)_{LB} \subset SO(6)_B$ and $SU(2)_D$ symmetries are gauged with the same $SU(2)_B$ gauge bosons.  The diagonal subgroup of $SU(2)_A\times SU(2)_B$ is then identified as the Standard Model $SU(2)_L$. Meanwhile, the diagonal combination of $SU(2)_{RA}\times SU(2)_{RB} \subset SO(6)_A\times SO(6)_B$ is gauged by the hypercharge $U(1)_Y$, while leaving the $\Delta$ sector unchanged.  This symmetry breaking leads to an extra heavy gauge boson triplet ($Z^\prime$, $W^{\prime\pm}$), with large squared masses proportional to $f^2+F^2$, which reduces their contribution to precision electroweak observables. Fermions in the BLH model, including the newly introduced top partners ($T$, $B$, $T_b^{2/3}$, $T_b^{5/3}$, $T_5$ and $T_6$), only transform under the global $SO(6)_A\times SO(6)_B$. This leads to top partners with masses proportional only to the scale $f$, lighter than the heavy gauge bosons, and results in a lesser degree of fine tuning than in other Little Higgs models \cite{Reuter:2012sd,Schmaltz:2010ac}. 

The BLH model has a large parameter space, allowing for a wide range of experimental signatures that could potentially reproduce either the CMS or the ATLAS results. Since the BLH model is a Type I two Higgs doublet model (2HDM), it also presents the possibility for a near-degeneracy between two or three physical scalar fields ($h_0$, $H_0$ and $A_0$), which would have a large effect on the measured signal rates of the observed scalar resonance. In particular, since a CP-odd scalar ($A_0$) boson does not couple directly to pairs of gauge bosons ($WW^*$ and $ZZ^*$), it is possible for a (nearly) degenerate CP-odd scalar to contribute to the diphoton rate without affecting these signal strengths. Type I 2HDM are not as strongly affected by meson factory constraints as Type II 2HDM \cite{Ferreira:2012nv}, and so the near-degenerate case presents a very interesting possibility that we explore in this paper. Although this scenario can lead to an enhancement in $\mu_{\gamma\gamma}$, it also leads to a large enhancement in $\mu_{\tau\tau}$.  As we will show, this effectively rules out the entirety of the near-degenerate scenario.

Alternatively, the large number of new vector boson, fermion and scalar fields in the BLH model can contribute to the loop-induced production and decay modes of a light Higgs boson. These additional states can also reproduce the observed enhancement of the diphoton rate without significantly affecting the non-loop-induced couplings \cite{Bonne:2012im}. Furthermore, mixing between flavour eigenstates in the BLH model further leads to modifications of the couplings from the normal SM expressions that can also result in changes to the Higgs boson signal strengths. These three features (extra Higgs states, new gauge and fermion states, modified couplings) combined lead to the possibility of large variances in the Higgs boson signal strength rates, and the potential to reproduce either the ATLAS or CMS measurements.

In this paper, we explore the Higgs results in the BLH model,  
accounting for all production and decay modes explored separately by ATLAS and CMS, in two scenarios: a general case, which assumes the $h_0$ state is light ($\approx 125$ GeV) and the masses of the other states ($H_0$ and $A_0$) are allowed to vary, and a second case with a near-degeneracy between the masses of the $h_0$ and $A_0$ fields. The masses of the $h_0$ and $A_0$ fields are input parameters for the model, while the $H_0$ mass is calculated from these input parameters and from the values of $\tan\beta$, the ratio of the vacuum expectation values of the two Higgs doublets, and $v$ (see Eq. \ref{derived_params}). Therefore, the $H_0$ state may or may not be similarly near-degenerate in the latter scenario, depending on the values of the input parameters ($m_{h_0}, m_{A_0}, \tan\beta, v$). These two regions are not orthogonal, as the general scenario does allow for the possibility of near degeneracy in the masses of the $A_0$ and $H_0$ states; this will be discussed further in Sec. \ref{sec:general_scenario}. In Sec. \ref{sec:calc}, we describe the formalism we use in our calculations of the Higgs results, while in Sec. \ref{sec:model} we describe the details of the BLH model that are relevant to our calculations. In Sec. \ref{sec:results}, we compare the BLH model predictions to the measured results from ATLAS and CMS in both scenarios. We summarize our results in Sec. \ref{sec:summary}.

\section{Calculations \label{sec:calc}}

\subsection{Production and Decay}

As shown in \cite{Gunion:1989we,Martinez:1989bg,Han:2003gf}, the scalar interactions of a Little Higgs model Lagrangian can be normalized to the form of the SM expressions by introducing scaling factors $y_i$, such that
\begin{widetext}
\begin{eqnarray}
\mathcal{L}_h &=& -\sum_f C_{S_0 \bar{f}f} S_0 \bar{f}f + \sum_V C_{S_0 VV} S_0 V^\dagger V - \sum_S C_{S_0 SS} S_0 S^\dagger S\ + \sum_{S^\prime,V} g_{S_0 S^\prime V}S_0 S^{\prime\dagger} V^{\mu} (p_0-p^{\prime})_\mu \cr\cr
&=& -\sum_f \frac{m_f}{v} y_{S_0 \bar{f}f} S_0 \bar{f}f + \sum_V 2 \frac{m_V^2}{v}y_{S_0 VV} S_0 V^\dagger V - \sum_S 2 \frac{m_S^2}{v}y_{S_0 SS} S_0 S^\dagger S + \sum_{S^\prime,V} g_{S_0 S^\prime V}S_0 S^{\prime\dagger} V^{\mu} (p_0-p^{\prime})_\mu
\label{eq:blhlag}
\end{eqnarray}
\end{widetext}
for fermion species $f$, vector bosons $V$ and scalars $S$. The $S_0$ label denotes the $h_0$, $H_0$, and $A_0$ (with an appropriate $\gamma_5$ factor) for the fermion interactions, and the $h_0$ and $H_0$ for the vector boson and scalar boson interaction terms. The parameters ($v$, $m_i$, $C_i$, $y_i$) are model dependent. The expression for the vev, $v$, will be discussed further in Sec. \ref{sec:model}.

\begin{figure}[tp]
   \begin{center}
	\includegraphics[width=0.5\textwidth]{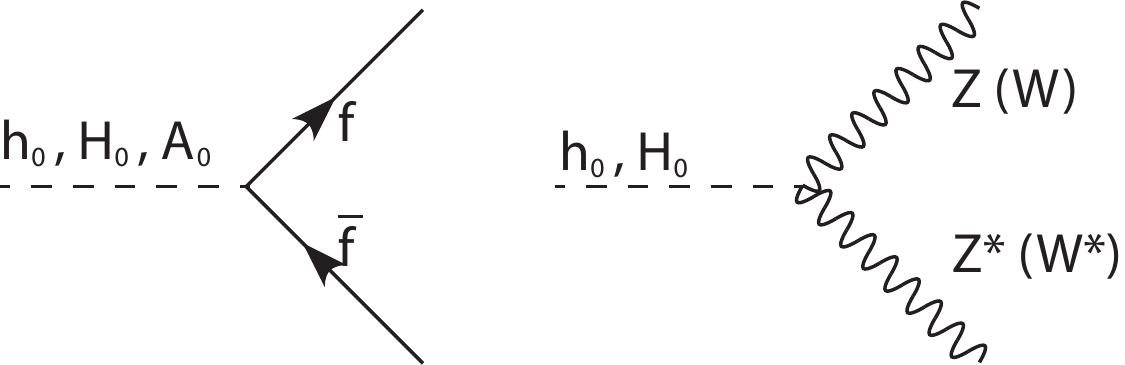}
   \end{center}
   \vspace{-4mm}
   \caption{Direct decay modes of the $h_0$, $H_0$ and $A_0$ states.}
\label{fig:partialdirect}
\end{figure}

\begin{figure*}[htbp]
   \begin{center}
	\includegraphics[width=0.9\textwidth]{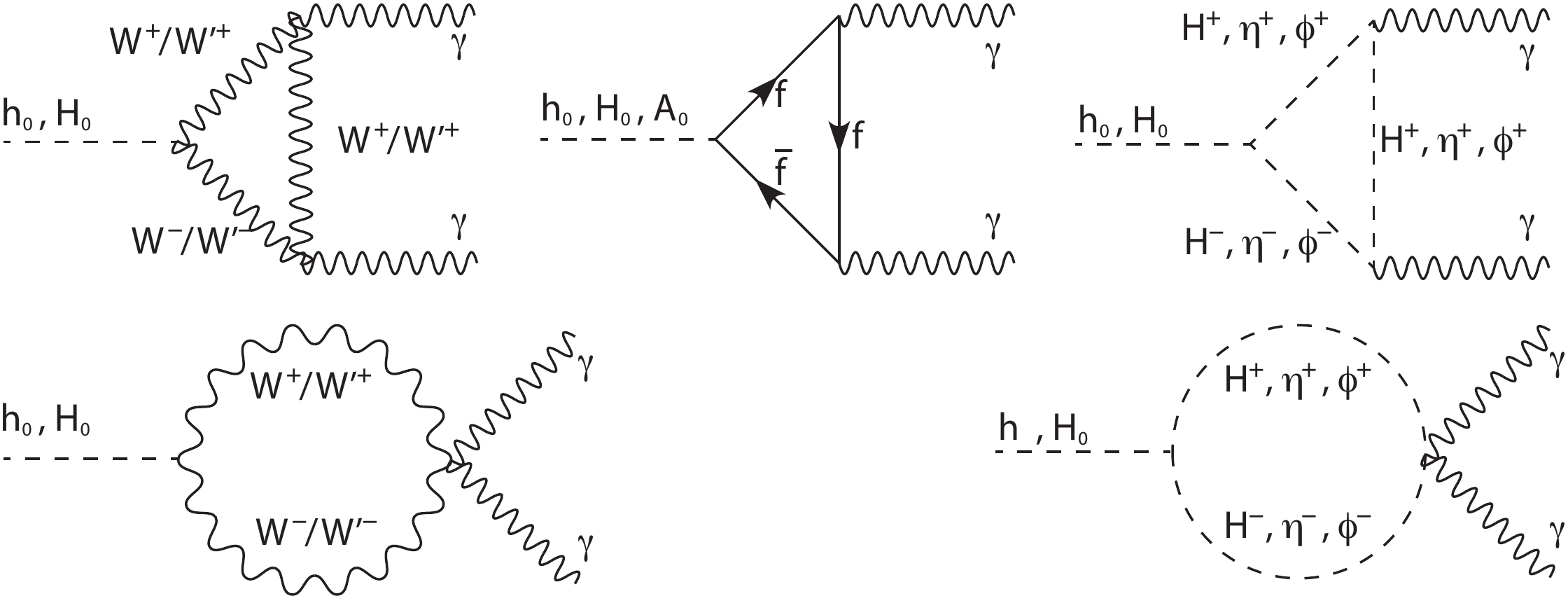}
   \end{center}
   \vspace{-4mm}
   \caption{Loop diagrams contributing to decays to pairs of photons, where $f$ refers to all fermions with an electric charge, including heavy, vector-like quark states. Similar diagrams exist for the $Z\gamma$ final state, where one of the photons is replaced by a $Z$ boson.}
\label{fig:partialyy}
\end{figure*}

\begin{figure}[htbp]
   \begin{center}
	\includegraphics[width=0.3\textwidth]{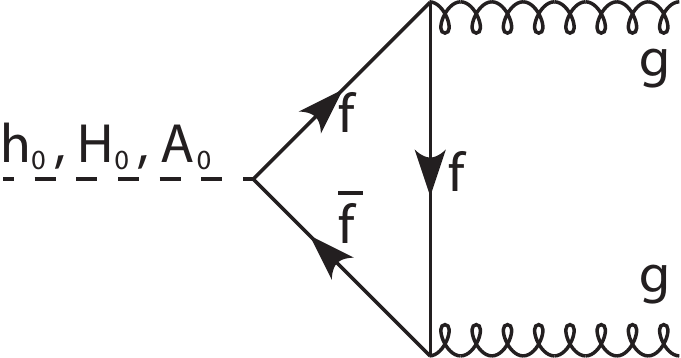}
   \end{center}
   \vspace{-4mm}
   \caption{Loop diagrams contributing to decays to pairs of gluons, where $f$ refers to all fermions with a colour charge, including heavy, vector-like quark states. This is also the process for the gluon-fusion production mode.}
\label{fig:partialgg}
\end{figure}

A Higgs boson of approximately 125 GeV decays predominantly to pair produced, kinematically accessible states, such as $b\bar{b}$ and $\tau^+\tau^-$, and to one on-shell and one off-shell vector boson ($ZZ^*$ and $WW^*$). Decays to diphotons, digluons and $Z\gamma$ also occur through loop interactions~\cite{Gunion:1989we}. In the BLH model, all three physical Higgs states ($h_0$, $H_0$ and $A_0$) can decay to light fermions and to loop-induced final states, but only the CP-even states ($h_0$ and $H_0$) can decay to pairs of weak gauge bosons, as shown in Figures \ref{fig:partialdirect}, \ref{fig:partialyy} and \ref{fig:partialgg}. In our calculations, we account for the contributions from all three Higgs states, and include all new particle states of the model that contribute to the loop diagrams.

When the scalar couplings are expressed in the form of Eq. \ref{eq:blhlag}, the partial decay widths of the $S_0=h_0,A_0,H_0$ can be written in terms of the SM calculated values (we use the values calculated in \cite{Dittmaier:2011ti}) multiplied by some combination of scaling factors. For direct decays, this is straightforward. For loop-induced decays the scaling factors must also include the loop factors. The direct decays are given by
\begin{eqnarray} \label{directdecays}
\Gamma(S_0 \rightarrow b\bar{b})_{BLH} &=& r_{S_0b\bar{b}}~\Gamma(S_0 \rightarrow b\bar{b})_{SM}  \cr\cr
\Gamma(S_0 \rightarrow c\bar{c})_{BLH} &=& r_{S_0c\bar{c}}~\Gamma(S_0 \rightarrow c\bar{c})_{SM} \cr\cr
\Gamma(S_0 \rightarrow \tau^+\tau^-)_{BLH} &=& r_{S_0\tau\tau}~\Gamma(S_0 \rightarrow \tau^+\tau^-)_{SM} \cr\cr
\Gamma(S_0 \rightarrow ZZ^*)_{BLH} &=& r_{S_0ZZ}~\Gamma(S_0 \rightarrow ZZ^*)_{SM} \cr\cr
\Gamma(S_0 \rightarrow WW^*)_{BLH} &=& r_{S_0WW}~\Gamma(S_0 \rightarrow WW^*)_{SM}
\end{eqnarray}
where
\begin{eqnarray} \label{directdecays2}
 r_{S_0b\bar{b}} &\equiv& \left(\displaystyle\frac{y_{S_0b\bar{b}}}{ y_{v}}\right)^2 \cr\cr
 r_{S_0c\bar{c}} &\equiv& \left(\displaystyle\frac{y_{S_0c\bar{c}}}{y_{v}}\right)^2 \cr\cr
 r_{S_0\tau\tau} &\equiv& \left(\displaystyle\frac{y_{S_0\tau\tau}}{y_{v}}\right)^2 \cr\cr
 r_{S_0ZZ} &\equiv& \left(\displaystyle\frac{y_{S_0ZZ} y_{Z\bar{f}f}^2}{y_v}\right)^2 \cr\cr
 r_{S_0WW} &\equiv& \left(\displaystyle\frac{y_{S_0WW} y_{W\bar{f}f^\prime}^2 y_{m_W}}{y_v}\right)^2.
\end{eqnarray}
The factor $y_v$ accounts for the differences between the vev in the BLH model and the SM vev, such that $v = y_v v_{SM}$, and is discussed in further detail in the following section.  The factor $y_{m_W} \equiv m_{W}^{BLH}/m_W^{SM}$ is the ratio of the $W$ boson mass calculated in the BLH model and in the SM, and appears when making the replacement $g=2m_W/v$ in the expression for the $S_0\rightarrow WW^*$ partial width \cite{Keung:1984hn}.  

The factors of $y_{V\bar{f}f}$ are given by $y_{W\bar{f}f^\prime} = (g_{W\bar{f}f^\prime}^{BLH})_L/(g_{W\bar{f}f^\prime}^{SM})_L$ and $y_{Z\bar{f}f} = \sqrt{((g_{Z\bar{f}f}^{BLH})_{L}^2+(g_{Z\bar{f}f}^{BLH})_{R}^2)/((g_{Z\bar{f}f}^{SM})_{L}^2+(g_{Z\bar{f}f}^{SM})_{R}^2)}$, and account for differences in the expressions of the couplings of the light vector bosons to fermions in the BLH model.  For our chosen parameter values, we found that these factors  amount to at most a 1\% correction to the $Z\bar{f}f$ and $W\bar{f}f^\prime$ couplings. Although these corrections are numerically small, it is appropriate to include them when the experimental measurements rely on leptonic decay modes of the vector bosons. Thus, we account for $Z\rightarrow l^+l^-$ and $W^\pm\rightarrow l^\pm \nu_l$ decays ($l=e,\mu$) in the calculation of the $ZZ^*$ and $WW^*$ decay widths, corresponding to the experimental results from ATLAS and CMS \cite{Cheung:2013kla,ATLAS:2012znl,ATLAS:2012klq,ATLAS:2012dsy,ATLAS:2012bmv,CMS:aya,CMS:yxa,CMS:lng}. Likewise, the $Z\rightarrow l^+l^-$ decay ($l=e,\mu$) is taken into account in the $Z \gamma$ channel as measured by ATLAS and CMS  \cite{ATLAS:2013Zy,CMS:2013Zy},  by including a factor of $y_{Z\bar{f}f}^2$ in the expression for $\Gamma(S_0 \rightarrow Z\gamma)_{BLH}$ in Equations \ref{eq:loopdec} and \ref{eq:loopdec_r} below.

The expressions for the partial widths of the loop-induced decay modes are given by
\begin{eqnarray}
\Gamma(S_0 \rightarrow \gamma\gamma)_{BLH} &=& r_{S_0\gamma\gamma}\Gamma(S_0\rightarrow \gamma\gamma)_{SM}\cr\cr
\Gamma(S_0 \rightarrow Z\gamma)_{BLH} &=& r_{S_0Z\gamma}\Gamma(S_0\rightarrow Z\gamma)_{SM}\cr\cr
\Gamma(S_0 \rightarrow gg)_{BLH} &=& r_{S_0gg}\Gamma(S_0\rightarrow gg)_{SM}
\label{eq:loopdec}
\end{eqnarray}
where
\begin{eqnarray}
r_{S_0\gamma\gamma}&\equiv& \displaystyle\frac{\left| \displaystyle\sum_{f_{SM}} A_{f,\gamma\gamma}^{BLH}+ A_{W,\gamma\gamma}^{BLH} + A_{new,\gamma\gamma}^{BLH}\right|^2}{y_v^2 \left| \displaystyle\sum_{f_{SM}} A_{f,\gamma\gamma}^{SM}+ A_{W,\gamma\gamma}^{SM}\right|^2} \cr\cr
r_{S_0Z\gamma} &\equiv& \displaystyle\frac{\left| \displaystyle\sum_{f_{SM}} A_{f,Z\gamma}^{BLH}+ A_{W,Z\gamma}^{BLH} + A_{new,Z\gamma}^{BLH}\right|^2}{y_v^2 \left| \displaystyle\sum_{f_{SM}} A_{f,Z\gamma}^{SM}+ A_{W,Z\gamma}^{SM}\right|^2} y_{Z\bar{f}f}^2 \cr\cr
r_{S_0gg} &\equiv& \displaystyle\frac{\left| \displaystyle\sum_{q_{SM}} A_{q,gg}^{BLH} + A_{new,gg}^{BLH}\right|^2}{y_v^2 \left| \displaystyle\sum_{q_{SM}} A_{q,gg}^{SM}\right|^2}
\label{eq:loopdec_r}
\end{eqnarray}

In this form, the $A_{new}^{BLH}$ terms in each channel account for the contributions from loops involving only the new (non-SM) particles of the BLH model. Thus, we find the expressions
\begin{eqnarray}
A_{new,\gamma\gamma}^{BLH} &=& \displaystyle\sum_f A_{f,\gamma\gamma}^{BLH} + A_{W^\prime,\gamma\gamma}^{BLH} + \displaystyle\sum_S  A_{S,\gamma\gamma}^{BLH}\cr\cr
A_{new,Z\gamma}^{BLH} &=& 2\displaystyle\sum_f A_{f,Z\gamma}^{BLH} + A_{W^\prime,Z\gamma}^{BLH} + 2\displaystyle\sum_S  A_{S,Z\gamma}^{BLH}\cr\cr
A_{new,gg}^{BLH} &=& \displaystyle\sum_f A_{f,gg}^{BLH}.
\label{eq:loopdec2}
\end{eqnarray}
The expressions for the $A$ terms in the BLH model are given by \cite{Djouadi:2005gi, Djouadi:2005gj}: 
\begin{eqnarray}
A_{f,\gamma\gamma}^{BLH} &\equiv&  y_{S_0\bar{f}f} \, Q_f^2 \, N_{c}^f \,F_{f}(\tau_f)\cr\cr
A_{V,\gamma\gamma}^{BLH} &\equiv& y_{S_0VV} \, Q_V^2  \, F_{V}(\tau_V)\cr\cr
A_{S,\gamma\gamma}^{BLH} &\equiv& y_{S_0SS} \, Q_S^2 \, F_{S}(\tau_S)\cr\cr
A_{f,Z\gamma}^{BLH} &\equiv& y_{S_0\bar{f}f} \left[(g_{Z\bar{f}f}^{BLH})_{L}+(g_{Z\bar{f}f}^{BLH})_{R}\right]  Q_f N_{c}^f F_{f}(\tau_f,\lambda_f)\cr\cr
A_{V,Z\gamma}^{BLH} &\equiv& y_{S_0 VV} \, g_{ZVV}^{BLH} \, Q_V \, F_{V}(\tau_V,\lambda_V)\cr\cr
A_{S,Z\gamma}^{BLH} &\equiv& y_{S_0 SS} \, g_{ZSS}^{BLH} \, Q_S \, F_{S}(\tau_S,\lambda_S)\cr\cr
A_{f,gg}^{BLH} &\equiv& y_{S_0\bar{f}f} \, F_{f}(\tau_f),
\label{eq:loopdec3}
\end{eqnarray}
where the SM expressions from the denominators in Eq. \ref{eq:loopdec_r} can be found by setting the $y_i$ scaling factors in Eq. \ref{eq:loopdec3} to unity. In these expressions, $Q$ is the electric charge of the particle, and $N_c$ is the number of colours (3 for quarks, 1 for leptons). The form factors $F_f(\tau_f)$, $F_V(\tau_V)$ and $F_S(\tau_S)$ (and similarly for $F_f(\tau_f,\lambda_f)$, $F_V(\tau_V,\lambda_V)$ and $F_S(\tau_S,\lambda_S)$) are found by integrating over the fermion ($f$), gauge boson ($V$) and scalar ($S$) loops, respectively. These are given by \cite{Djouadi:2005gi, Djouadi:2005gj}: 
\begin{eqnarray}
F_f(\tau) &=& \left\{ \begin{array}{l l} 2\tau(1+(1-\tau)f(\tau)) & \textrm{for } S_0=h_0, H_0 \\
                            2\tau f(\tau) & \textrm{for } S_0=A_0
                \end{array} \right. \cr\cr
F_V(\tau) &=& -(2 + 3\tau +3\tau(2-\tau)f(\tau))\cr\cr
F_S(\tau) &=& -\tau(1-\tau f(\tau))\cr\cr
F_f(\tau, \lambda) &=& \left\{ \begin{array}{l l} I_1(\tau,\lambda) - I_2(\tau,\lambda)& \textrm{for } S_0=h_0, H_0 \\
							- I_2(\tau,\lambda) & \textrm{for } S_0=A_0 
				\end{array} \right. \cr\cr 
F_V(\tau,\lambda) &=& 4(3-\tan^2\theta_w)I_2(\tau,\lambda) \nonumber \\
&& +((1+2\tau^{-1})\tan^2\theta_w-5-2\tau^{-1})I_1(\tau,\lambda)\cr\cr
F_S(\tau,\lambda) &=& -I_1(\tau,\lambda)
\end{eqnarray}
where
\begin{eqnarray}
I_1(x,y)&=&\frac{xy}{2(x-y)}+\frac{x^2y^2}{2(x-y)^2}(f(x)-f(y)) \nonumber \\  && \qquad \qquad\, + \frac{x^2y}{(x-y)^2}(g(x)-g(y)) \cr\cr
I_2(x,y)&=& -\frac{xy}{2(x-y)}(f(x)-f(y)) \cr\cr\cr
f(x) &=& \displaystyle\left\{\begin{array}{l l}\left[\arcsin(1/\sqrt{x})\right]^2, & x \ge 1\\
-\displaystyle\frac{1}{4}\left[\ln\left(\frac{1+\sqrt{1-x}}{1-\sqrt{1-x}}\right) -i \pi\right]^2, & x < 1\end{array}\right. \cr\cr\cr
g(x) &=& \displaystyle\left\{\begin{array}{l l}\sqrt{x-1}\arcsin(1/\sqrt{x}), & x \ge 1\\
\displaystyle\frac{\sqrt{1-x}}{2}\left[\ln\left(\frac{1+\sqrt{1-x}}{1-\sqrt{1-x}}\right) -i \pi\right], & x < 1\end{array}\right.   \nonumber \\
\end{eqnarray} 
In the above, $\tau_{f,V,S} \equiv 4m_{f,V,S}^2/m_{S_0}^2$ and $\lambda_{f,V,S} \equiv 4m_{f,V,S}^2/m_{Z}^2$, for field $f$, $V$ or $S$, with corresponding mass $m_{f,V,S}$. In the limit that $\tau_i \gg 1$, these factors approach the values of $F_f \rightarrow 4/3$, $F_V \rightarrow -7$, and $F_S \rightarrow 1/3$.

In the BLH model, it is useful to consider the contributions of each of three sectors - gauge, fermion and scalar - to the diphoton rate. This can be performed by summing up all contributing states $S_{h\gamma\gamma-G}=\sum A_{V,\gamma\gamma}^{BLH}$, $S_{h\gamma\gamma-F}=\sum A_{f,\gamma\gamma}^{BLH}$ and $S_{h\gamma\gamma-S}=\sum A_{S,\gamma\gamma}^{BLH}$. The relative value and sign of these terms, $S_{h\gamma\gamma-G}$, $S_{h\gamma\gamma-F}$, and $S_{h\gamma\gamma-S}$, represent the size of the contribution of that sector of particle states to the diphoton effective coupling. The diphoton effective coupling is dominated by gauge boson loops in the SM, with a sub-dominant contribution from fermions (predominantly the top quark). This is not expected to change significantly in the BLH model in regions of parameter space that produce SM-like signal strength ratios. The reason for this is that, for example, a decrease in the gauge contribution to the diphoton rate would likely correspond to a smaller $W^+W^-$ rate, assuming a negligible contribution from the heavy $W^\prime$.

\begin{figure*}[htbp]
   \begin{center}
	\includegraphics[width=0.7\textwidth]{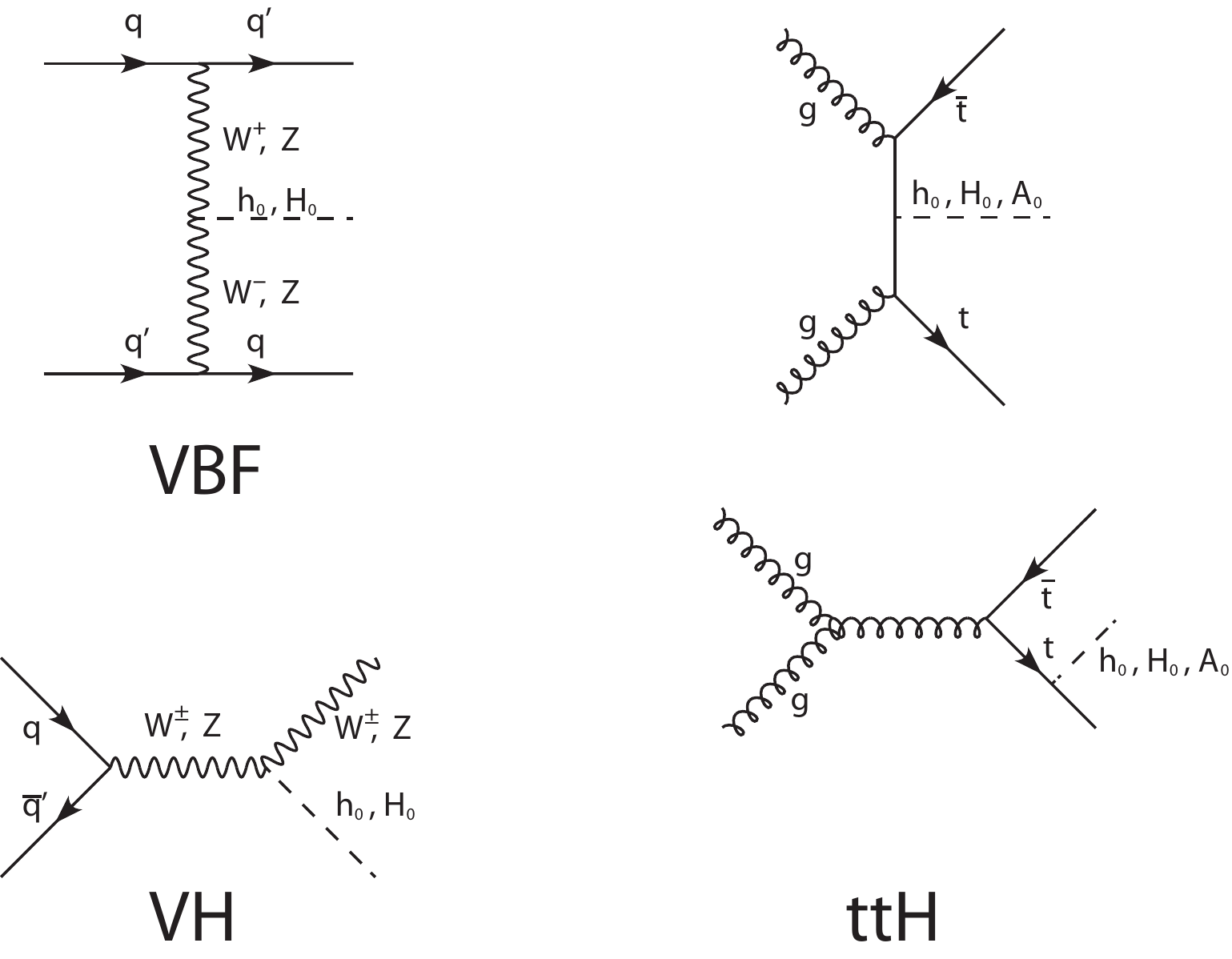}
   \end{center}
   \caption{Subdominant production modes for Higgs bosons at the LHC - vector boson fusion (VBF), associated production with a vector boson (VH), and associated production with top quarks (ttH).}
\label{fig:production}
\end{figure*}

For new fields whose masses are proportional to $f$ or $F$, such as the new vector-like quarks and heavy gauge bosons in the BLH model, the scaling factors behave as $y_i \propto (v/f)^k$, where $k\ge 1$. Thus, while the factor of $F_i(\tau)$ may increase in magnitude with increasing mass of the new state, the scaling factor, $y_i$, decreases at a faster rate, and very high mass states typically have only a small contribution to the loop factors.  It is also possible that loops involving both $W^\pm$ and $W^{\prime\pm}$ or $\phi^\pm$ and $\eta^\pm$ contribute to the $S_0\rightarrow\gamma\gamma$ and $S_0\rightarrow Z\gamma$ decay widths due to vertices of the form of $h W^+ W^{\prime -}$ and $h \phi^+ \eta^-$.  However, since the $\gamma W^+ W^{\prime -}$ and $\gamma\phi^+\eta^-$ couplings of the BLH model are suppressed at $\mathcal{O}(v^2/f^2)$, these mixed loops can be safely neglected in this model.

In addition to the decays discussed above, the heavy CP-even scalar, $H_0$, and the CP-odd scalar, $A_0$, of the BLH model, when kinematically allowed, may also decay to top quarks ($H_0 ,A_0 \to t\bar{t}^{(*)}$), pairs of scalars ($H_0 \to hh^{(*)}, A_0A_0^{(*)}$, $H^+ H^{-(*)}$), or a scalar and a gauge boson ($H_0 \rightarrow A_0 Z^{(*)}, H^\pm W^{\mp(*)}$ and $A_0 \rightarrow h Z^{(*)}$, where in the case of off-shell decays, $Z^*\to f\bar{f}$ and $W^* \to f\bar{f}^\prime$) \cite{Djouadi:1996}. These decays have the effect of lowering the $H_0$ and $A_0$ branching ratios to the standard modes considered above. Hence, their contributions to the Higgs signal strengths under consideration in this study are lowered, particularly for $H_0$ and $A_0$ masses above roughly 200-300~GeV \cite{Djouadi:1996}. Thus, it is important to include their effects in our analysis.

The partial widths for the top quark decays of $S_0 = H_0, A_0$ were calculated using the data from \cite{Dittmaier:2011ti}, scaled in a similar manner as the direct decays of Equations \ref{directdecays} and \ref{directdecays2}:
\begin{eqnarray}
\Gamma(S_0 \rightarrow t\bar{t})_{BLH} =&  \beta_t^p \left(\displaystyle\frac{y_{S_0t\bar{t}}}{ y_{v}}\right)^2 \Gamma(S_0 \rightarrow t\bar{t})_{SM}
\end{eqnarray}
where $\beta_t \equiv \sqrt{1- 4m_t^2/m_{S_0}^2}$ and $p = 0 (-2)$ for $S_0 = H_0 (A_0)$ is an additional factor that accounts for the difference in kinematics of the CP-even $H_0$ and CP-odd $A_0$ decays.

The scalar-scalar and scalar-gauge partial widths were calculated using HDECAY \cite{Djouadi:1998} in a generic 2HDM using the mixing angle parameters $\alpha_{2HDM}=-0.14$ and $\tan\beta_{2HDM}=1.5$ for a range of scalar masses from 125~GeV to 1200~GeV in 5~GeV increments.  To calculate the widths for a particular set of BLH model parameters, interpolation was used on the discrete set of 2HDM widths and the results were scaled according to
\begin{eqnarray}
\Gamma(H_0 \rightarrow A_0Z)_{BLH} &=& r_{HAZ} ~\Gamma(H_0 \rightarrow A_0Z)_{2HDM}  \cr\cr
\Gamma(H_0 \rightarrow H^\pm W^\mp )_{BLH} &=& r_{HHW} ~\Gamma(H_0 \rightarrow H^\pm W^\mp)_{2HDM}  \cr\cr
\Gamma(H_0 \rightarrow hh)_{BLH} &=& r_{Hhh}  ~\Gamma(H_0 \rightarrow hh)_{2HDM}  \cr\cr
\Gamma(H_0 \rightarrow A_0A_0)_{BLH} &=& r_{HAA} ~\Gamma(H_0 \rightarrow A_0A_0)_{2HDM} \cr\cr
\Gamma(H_0 \rightarrow H^+ H^-)_{BLH} &=& r_{HHH}  ~\Gamma(H_0 \rightarrow H^+ H^-)_{2HDM} \cr\cr
\Gamma(A_0 \rightarrow hZ)_{BLH} &=& r_{AhZ}  ~\Gamma(A_0 \rightarrow hZ)_{2HDM}
\end{eqnarray}
where
\begin{eqnarray}
r_{HAZ} &\equiv& \left(\frac{(g_{H_0A_0Z})_{BLH}}{(g_{H_0A_0Z})_{2HDM}}\right)^2  \cr\cr
r_{HHW} &\equiv& \left(\frac{(g_{H_0 H^\pm W^\mp})_{BLH}}{(g_{H^0 H^\pm W^\mp})_{2HDM}}\right)^2  \cr\cr
r_{Hhh} &\equiv& \left(\frac{(g_{H_0hh})_{BLH}}{(g_{H_0hh})_{2HDM}}\right)^2  \cr\cr
r_{HAA}  &\equiv& \left(\frac{(g_{H_0A_0A_0})_{BLH}}{(g_{H_0A_0A_0})_{2HDM}}\right)^2   \cr\cr
r_{HHH} &\equiv& \left(\frac{(g_{H_0H^+H^-})_{BLH}}{(g_{H_0H^+H^-})_{2HDM}}\right)^2 \cr\cr
r_{AhZ} &\equiv& \left(\frac{(g_{A_0hZ})_{BLH}}{(g_{A_0hZ})_{2HDM}}\right)^2
\end{eqnarray}
where the expressions for the 2HDM couplings, $g_{2HDM}$, used in HDECAY are written in Appendix E of \cite{Kanemura:2004}.  We have verified that these couplings agree with those listed in Appendix A of \cite{Gunion:1989we}.  The couplings, $g_{BLH}$, of the BLH model are directly related to the scaling factors, $y_i$, and are listed in Eq. \ref{y_SSS} and \ref{g_SSV} of Section \ref{sec:model}.  

Expressions similar to those for the decay modes exist for the production modes. At the LHC, the primary production mode is through gluon-gluon fusion (ggF), which occurs via the reverse of the diagram in Figure~\ref{fig:partialgg}. The other production mechanisms at the LHC include vector boson fusion (VBF), associated production with a vector boson (VH), and associated production with top quarks (ttH). The diagrams for these subdominant processes are given in Fig.~\ref{fig:production}.

In the BLH model, the cross sections are modified from the SM expressions in a similar manner as the decay widths, and can be expressed as a multiplicative factor times the SM values, such as:
\begin{eqnarray}
\sigma(ggF)_{BLH} &=& r_{S_0gg} \, \sigma(ggF)_{SM}\cr\cr
\sigma(VBF)_{BLH} &=& (y_{S_0WW}^2 \, y_{m_W}^2/y_v^2) \, \sigma(VBF)_{SM}\cr\cr
\sigma(WH)_{BLH} &=& (y_{S_0WW}^2 \, y_{m_W}^2/y_v^2) \, \sigma(WH)_{SM}\cr\cr
\sigma(ZH)_{BLH} &=& (y_{S_0ZZ}^2/y_v^2) \, \sigma(ZH)_{SM}\cr\cr
\sigma(VH)_{BLH} &=& \sigma(WH)_{BLH} +\sigma(ZH)_{BLH} \cr\cr
\sigma(ttH)_{BLH} &=& y_{S_0 t\bar{t}}^2 \, \sigma(ttH)_{SM}
\end{eqnarray}
The factor modifying the gluon fusion production mode, $r_{S_0gg}$, is the same as for the $gg$ decay mode, defined in Eq. \ref{eq:loopdec_r}. Since $y_{m_W} \approx 1$ and $y_{S_0WW} = y_{S_0ZZ}$ in the BLH model due to custodial symmetry, the VBF and VH expressions are essentially independent of whether the gauge boson is a $Z$ or a $W$.

\section{Model Details and Couplings \label{sec:model}}

The following subsections deal with the determination of the scaling factors, $y_i$, for all neutral Higgs states ($h_0, H_0, A_0$) coupling to the charged scalars ($H^\pm, \phi^\pm, \eta^\pm$), the gauge bosons ($Z, W^\pm, W^{\prime\pm}$), the SM fermions, and the additional fermions ($T, B, T_b^{2/3}, T_b^{5/3}, T_5, T_6$) of the BLH model.

\subsection{Scalar Sector}

The BLH model Higgs fields, $h_1$ and $h_2$, form a two Higgs doublet potential that undergoes spontaneous symmetry breaking. The most basic form of this potential is given below as in \cite{Schmaltz:2010ac}, and is sufficient to understand EWSB in the model:
\begin{equation}
V = \frac{1}{2}m_1^2 h_1^T h_1 + \frac{1}{2}m_2^2 h_2^T h_2 - B_\mu h_1^T h_2 + \frac{\lambda_0}{2} (h_1^T h_2)^2.
\end{equation}
In this form, each of the Higgs ``doublets" are written as \textbf{4}'s of $SO(4)$, which have the same degrees of freedom and hypercharge values as the standard two Higgs doublets from \cite{Eriksson:2009ws}. Additionally, while there are other quartic terms present in the BLH model (namely $(h_1^T h_1)^2$, $(h_2^T h_2)^2$, $(h_1^T h_1)(h_2^T h_2)$, and $(h_1^T h_1+h_2^T h_2)(h_1^T h_2)$), their coefficients are generated at loop level and do not significantly affect the details of electroweak symmetry breaking.
As these terms have a small effect on the diagonalization, they are ignored in the determination of the mass eigenstates. They are not ignored in the determination of the couplings, however, as they comprise the dominant contribution to the interactions of the scalar triplets with the Higgs fields.

The parameters in this potential are generated in part explicitly and in part radiatively, as with other Little Higgs models, and spontaneous symmetry breaking occurs for parameter values that satisfy $B_\mu > m_1 m_2$. The minimization condition is achieved by shifting the first component of each of $h_1$ and $h_2$ by a respective vacuum expectation value (vev), $v_1$ and $v_2$. These vevs can be parameterized by a mixing angle, $\tan\beta=v_1/v_2 = m_2/m_1$, where $v=\sqrt{v_1^2+v_2^2}$ is related to the Standard Model vev via the relation $v=y_v v_{SM}$ (the expression for $y_v$ will be addressed in the next subsection). 

Diagonalizing the mass matrix for the scalar sector results in three physical neutral scalar fields ($h_0$, $H_0$ and $A_0$, along with the unphysical $G_0$), and two physical charged scalar fields ($H^\pm$, along with the unphysical $G^\pm$), parameterized by an angle $\alpha$ such that
\begin{eqnarray}
h_{1}[1] &=& \cos\alpha \, h_0 - \sin\alpha \, H_0 + v \sin\beta \nonumber\\
h_{2}[1] &=& \sin\alpha \, h_0 + \cos\alpha \, H_0 + v \cos\beta \nonumber\\
h_{1}[2] &=& -\cos\beta \, A_0 + \sin\beta \, G_0 \nonumber\\
h_{2}[2] &=& +\sin\beta \, A_0 + \cos\beta \, G_0 \nonumber\\
h_{1}^\pm &=& -\cos\beta \, H^\pm + \sin\beta \, G^\pm \nonumber\\
h_{2}^\pm &=& +\sin\beta \, H^\pm + \cos\beta \, G^\pm .
\end{eqnarray}

The four parameters in the Higgs potential ($m_1$, $m_2$, $B_\mu$ and $\lambda_0$) can be replaced by a more phenomenologically accessible set consisting of the masses of the $h_0$ and $A_0$ states, along with the mixing angle $\beta$ and the vev, $v$.  In this parameterization, we find the following expressions:
\begin{widetext}
\begin{eqnarray}
\label{derived_params}
\lambda_0 &=& \frac{m_{h_0}^2}{v^2}\left( \frac{m_{h_0}^2-m_{A_0}^2}{m_{h_0}^2-\sin^2(2\beta) m_{A_0}^2} \right) \nonumber\\
B_\mu &=& \frac{1}{2}(m_{A_0}^2 + v^2 \lambda_0) \sin(2\beta) \nonumber\\
\tan\alpha &=& \frac{B_\mu \cot(2\beta) + \sqrt{B_\mu^2/\sin^2(2\beta) - 2 \lambda_0 B_\mu v^2 \sin(2\beta) + \lambda_0^2 v^4 \sin^2(2\beta)} }{B_\mu - \lambda_0 v^2 \sin(2\beta)} \nonumber\\
m_{H^\pm}^2 &=& m_{A_0}^2  = m_1^2 + m_2^2 \nonumber\\
m_{H_0}^2 &=& \frac{B_\mu}{\sin(2\beta)} + \sqrt{\frac{B_\mu^2}{\sin^2(2\beta)}-2\lambda_0 B_\mu v^2 \sin(2\beta) + \lambda_0^2 v^4 \sin^2(2\beta)} \nonumber\\
m_{\sigma}^2 &=& (\lambda_{56}+\lambda_{65}) f^2 \equiv 2 \lambda_0 f^2 K_\sigma ~ . 
\end{eqnarray}
\end{widetext}
Since the quartic interaction, with coupling $\lambda_0$, is produced when the heavy singlet state ($\sigma$) is integrated out, it is related to the fundamental parameters $\lambda_{56}$ and $\lambda_{65}$ via $\lambda_0=2\lambda_{56} \lambda_{65}/(\lambda_{65}+\lambda_{56})$. The parameters $\lambda_{56}$ and $\lambda_{65}$ are the coefficients of the quartic potential, defined in Eq. 9 of \cite{Schmaltz:2010ac}, and must both be non-zero to achieve collective symmetry breaking and generate a Higgs quartic coupling. Rather than expressing $m_\sigma^2$ in terms of these two free parameters, we instead choose to parameterize it in terms of $\lambda_0$ and a single free parameter, $K_\sigma$, as shown in the last line of Eq. \ref{derived_params}.

There exist a number of theoretical constraints that can be placed on these parameters, primarily due to perturbativity requirements. The value of the mixing angle $\beta$ is limited by two constraints, the first of which is the requirement that $\lambda_0 < 4\pi$, leading to an upper bound of
\begin{equation}
\tan\beta < \sqrt{\frac{2+2\sqrt{\left(1-\frac{m_{h_0}^2}{m_{A_0}^2}\right) \left(1-\frac{m_{h_0}^2}{4\pi v^2}\right)} }{\frac{m_{h_0}^2}{m_{A_0}^2}\left(1+\frac{m_{A_0}^2-m_{h_0}^2}{4\pi v^2}\right)} - 1}
\label{eq:tBmax}
\end{equation}
A lower bound also exists, and is set by examining the radiatively-induced contributions to $m_1$ and $m_2$ in the model, which suggest that $\tan\beta \gtrsim 1$~\cite{Schmaltz:2010ac}. Furthermore, there are limits on $K_\sigma$ from requiring that $\lambda_{56/65}$ are real valued and $< 4\pi$, such that
\begin{eqnarray}
\lambda_{56/65}&=&\frac{1}{2f^2}(m_\sigma^2 \pm \sqrt{m_\sigma^4 - 2f^2 m_\sigma^2 \lambda_0})~,
\end{eqnarray}
which leads to the bounds
\begin{eqnarray}
1 < K_\sigma < \frac{16\pi^2}{\lambda_0(8\pi - \lambda_0)}~.
\label{eq:sigmamax}
\end{eqnarray}

The BLH model also contains two physical, real triplets, with charged fields $\phi^\pm$ and $\eta^\pm$ that are relevant to the diphoton and $Z\gamma$ decay modes. These scalar triplet fields obtain a contribution to their mass from the explicit symmetry breaking terms in the model, as defined in Eq. 38 of \cite{Schmaltz:2010ac}, that depend on the parameter $m_4 \sim 10$~GeV. However, their masses are dominated by contributions from the Coleman-Weinberg potential involving gauge boson loops for $\phi^\pm$ and hypercharge boson loops for $\eta^\pm$. The dominant contributions 
 to their masses are given by
\begin{eqnarray}
m_{\phi^\pm}^2 &\approx& \kappa_G \frac{3}{32\pi^2} g_A^2 g_B^2\left(1-\frac{v^2}{2f^2}\frac{F^2}{f^2+F^2}\right)(f^2+F^2) \cr\cr 
&&\times \log\left(\frac{\Lambda^2}{m_{W^\prime}^2}\right)  \cr\cr
m_{\eta^\pm}^2 &\approx& \kappa_Y \frac{3}{16\pi^2}g^{\prime 2} \left(1-\frac{v^2}{2f^2} \right)\Lambda^2 ,
\end{eqnarray}
where $\Lambda \approx 4\pi f$ is the compositeness scale, and $\kappa_G$ and $\kappa_Y$ are taken as $\mathcal{O}(1)$ factors that account for the details of the cancellation of the gauge logarithmic, and hypercharge quadratic divergent loops, respectively, at the scale $\Lambda$. In our calculations, we take into account all contributions up to $\mathcal{O}(v^2/f^2)$, including the subdominant explicit symmetry breaking mass terms \cite{Schmaltz:2010ac}, but do not show them here for brevity. The factors $g_A$ and $g_B$ are the gauge couplings for $SU(2)_A$ and $SU(2)_B$, respectively, which will be discussed in further detail in Section \ref{sec:gauge}. 

The scaling factors for the charged scalar interactions are given by
\begin{widetext}
\begin{eqnarray} \label{y_SSS}
y_{h_0\eta^+\eta^-} &\approx& -(c_\beta s_\alpha + c_\alpha s_\beta)\frac{v^2}{2 f^2} = - s_{\alpha+\beta} \frac{v^2}{2 f^2}  \cr\cr
y_{H_0\eta^+\eta^-} &\approx& -(c_\beta c_\alpha - s_\alpha s_\beta)\frac{v^2}{2 f^2} = - c_{\alpha+\beta} \frac{v^2}{2 f^2} \cr\cr
y_{h_0\phi^+\phi^-} &\approx& -(c_\beta s_\alpha + c_\alpha s_\beta)\frac{v^2 F^2}{2 f^2 (f^2+F^2)} = - s_{\alpha+\beta} \frac{v^2 F^2}{2 f^2 (f^2+F^2)}\cr\cr
y_{H_0\phi^+\phi^-} &\approx& -(c_\beta c_\alpha - s_\alpha s_\beta)\frac{v^2 F^2}{2 f^2 (f^2+F^2)} = -c_{\alpha+\beta} \frac{v^2 F^2}{2 f^2 (f^2+F^2)}\cr\cr
y_{h_0H^+H^-} &\approx& \frac{v^2}{(768 f^2 m_{A_0}^2 \pi^2)}
\biggl(-9 \kappa_G F^2 g_A^2 g_B^2 \log\left[\frac{\Lambda^2}{m_{W^\prime}^2}\right] (c_\beta s_\alpha + c_\alpha s_\beta) - 32 \kappa_S \lambda_0 m_\sigma^2  \log\left[\frac{\Lambda^2}{m_\sigma^2}\right] (3 c_\beta s_\alpha + 2 c_\alpha s_\beta) \cr
&& + 128 \pi^2 c_\beta s_\beta (-6 f^2 \lambda_0 (c_\alpha c_\beta +s_\alpha s_\beta) + m_{A_0}^2 (c_\alpha c_\beta - s_\alpha s_\beta) (c_\beta^2 - s_\beta^2))\biggr)\cr\cr
y_{H_0H^+H^-} &\approx& \frac{v^2}{(768 f^2 m_{A_0}^2 \pi^2)}
\biggl(9 \kappa_G F^2 g_A^2 g_B^2 \log\left[\frac{\Lambda^2}{m_{W^\prime}^2}\right] (c_\beta c_\alpha - s_\alpha s_\beta) + 32 \kappa_S \lambda_0 m_\sigma^2  \log\left[\frac{\Lambda^2}{m_\sigma^2}\right] (3 c_\beta s_\alpha + 2 c_\alpha s_\beta)(c_\beta^2 - s_\beta^2) \cr
&& + 128 \pi^2 c_\beta s_\beta (-6 f^2 \lambda_0 (c_\beta s_\alpha - c_\alpha s_\beta) + m_{A_0}^2 (c_\beta s_\alpha + c_\alpha s_\beta) (c_\beta^2 - s_\beta^2))\biggr)\cr\cr
y_{H_0 A_0 A_0} &\approx& \frac{v^2}{(768 f^2 m_{A_0}^2 \pi^2)}
\biggl(9 \kappa_G F^2 g_A^2 g_B^2 \log\left[\frac{\Lambda^2}{m_{W^\prime}^2}\right] (c_\beta c_\alpha - s_\alpha s_\beta) + 32 \kappa_S \lambda_0 m_\sigma^2  \log\left[\frac{\Lambda^2}{m_\sigma^2}\right] (3 c_\beta s_\alpha + 2 c_\alpha s_\beta)(c_\beta^2 - s_\beta^2) \cr
&& + 36 \kappa_Y g^2 \tan^2\theta_w \Lambda^2 (c_\beta c_\alpha - s_\beta s_\alpha) + 128 \pi^2 c_\beta s_\beta (-6 f^2 \lambda_0 (c_\beta s_\alpha - c_\alpha s_\beta) + m_{A_0}^2 (c_\beta s_\alpha + c_\alpha s_\beta) (c_\beta^2 - s_\beta^2))\biggr)\cr\cr
y_{H_0 h_0 h_0} &\approx& \frac{v^2}{(768 f^2 m_{h}^2 \pi^2)} \biggl(9 \kappa_G F^2 g^4 \cot^2\theta_g \log\left[\frac{\Lambda^2}{m_{W^\prime}^2}\right] (c_\beta c_\alpha - s_\alpha s_\beta)\cr
&& + 96 \kappa_S \lambda_0 m_\sigma^2  \log\left[\frac{\Lambda^2}{m_\sigma^2}\right] ( c_\beta c_\alpha -s_\beta s_\alpha +s_\alpha c_\alpha (s_\beta c_\alpha -c_\beta s_\alpha ) ) + 12 \kappa_Y g^2 \tan^2\theta_w \Lambda^2 (c_\beta c_\alpha - s_\beta s_\alpha) \cr
&&  -3072 \pi^2 c_\beta s_\beta (f^2 \lambda_0 (c_\beta c_\alpha -s_\beta s_\alpha +3s_\alpha c_\alpha (s_\beta c_\alpha -c_\beta s_\alpha ) ) + m_{A_0}^2 ( (c_\alpha^2-s_\alpha^2)(c_\beta s_\alpha + s_\beta c_\alpha) )) \biggr) 
\end{eqnarray}
\end{widetext}
where $c_{\theta} = \cos\theta$, $s_{\theta} = \sin\theta$, and $t_\theta = \tan\theta$, for $\theta = \beta$ and $\alpha+\beta$.  The factor of $\kappa_S$ is also taken as an $\mathcal{O}(1)$ factor that accounts for the details of the cancellation of the logarithmic divergence involving scalar loops. Here we have also neglected contributions from higher orders in the expansion of $v/f$, and terms proportional to the small explicit symmetry breaking parameters in the BLH model \cite{Schmaltz:2010ac}.

\subsection{Gauge Sector \label{sec:gauge}}

The gauge couplings $g_A$ and $g_B$, associated with $SU(2)_A \times SU(2)_B$, can be parameterized in a more phenomenological fashion in terms of a mixing angle $\theta_g$ ($\tan\theta_g \equiv g_A / g_B$) and the $SU(2)_L$ gauge coupling, $g=g_A g_B / \sqrt{g_A^2 + g_B^2}$.  Furthermore, $g$ and $g^\prime$, the gauge coupling associated with the $U(1)_Y$ symmetry,  can be parameterized as in the SM in terms of the fundamental charge, $e$, and the weak mixing angle, $\theta_w$, such that $\sin\theta_w = g^\prime/\sqrt{g^2+g^{\prime2}}$, and $e=g\sin\theta_w$.

The masses of the gauge bosons in the BLH model are given by:
\begin{eqnarray}
m_W^2 &=& \frac{g^2 v^2}{4} y_{m_W}^2 \cr\cr
m_Z^2 &=& \frac{g^2 v^2}{4 c_w^2} y_{m_W}^2 \cr\cr
m_{W^\prime}^2 &=& \frac{g^2}{4 c_g^2 s_g^2}(f^2+F^2) - m_W^2\cr\cr
m_{Z^\prime}^2 &=& m_{W^\prime}^2 + \frac{g^2 s_w^2 v^4}{16 c_w^2 (f^2 + F^2)}(c_g^2-s_g^2)^2
\label{gauge_masses}
\end{eqnarray}
where 
\begin{equation}
y_{m_W}^2=1 - \frac{v^2}{6f^2}\left(1 + \frac{3}{2}\frac{f^2}{f^2+F^2}(c_g^2 - s_g^2)^2\right)
\end{equation}
and $c_g = \cos\theta_g$ and $s_g = \sin\theta_g$. 
While both $m_W^2$ and $m_Z^2$ obtain corrections of $\mathcal{O}(v^4/f^2)$, which do not vanish as $F \rightarrow \infty$, the model retains custodial symmetry and such factors cancel out in their contribution to the $\rho$ parameter.  Additionally, corrections to the $Z\bar{f} f$ couplings are proportional to $v^2/(f^2+F^2)$, and do vanish for $F \rightarrow \infty$ \cite{Schmaltz:2010ac}.

We use the experimental value of the fine structure constant, $\alpha_{EM}$, to determine the value of $e$ ($=\sqrt{\alpha_{EM}/4\pi}$), and calculate $v$ and $\sin\theta_w$ from the experimental values of  $G_F$ and the pole mass of the $Z$ boson, by exploiting the custodial symmetry relation, $\rho = 1$. Accounting for the contributions from the $W$ and $W^\prime$ in the process $\mu \rightarrow e \bar{\nu}_e \nu_\mu$, gives an expression for the vev, $v$, in terms of the SM vev, such that
\begin{equation}
v = y_v ~v_{SM}
\end{equation}
where
\begin{equation}
y_v =  1 + \frac{v_{SM}^2}{12f^2} \left(1 + \frac{f^2}{2 (f^2 + F^2)} (3 + 12 (c_g^2 - s_g^2)) \right). \\
\end{equation}
Combining this with the expression for the $Z$ pole mass ($\widehat{m}_Z$), the weak mixing angle in the BLH model can be expressed as
\begin{equation}
\sin^2\theta_w = \frac{1}{2}\left(1-\sqrt{1 - \frac{e^2 v^2}{\widehat{m}_Z^2} y_{m_W}^2 } \right)~.
\end{equation}

The scaling factors important to the VBF and VH production modes, and to the diboson decay modes ($S_0\rightarrow \gamma\gamma, VV^*$) are given by
\begin{eqnarray}
y_{h_0WW} = y_{h_0ZZ} &=& s_{\alpha+\beta} ~y_{m_W}^2 \cr\cr\cr
y_{H_0WW} = y_{H_0ZZ} &=& \frac{c_{\alpha+\beta}}{s_{\alpha+\beta}}~y_{h_0WW}  \cr\cr\cr
y_{h_0W^\prime W^\prime} &=&  -s_{\alpha+\beta} \frac{c_g^2 s_g^2 v^2}{f^2+F^2}\cr\cr\cr
y_{H_0W^\prime W^\prime} &=& \frac{c_{\alpha+\beta}}{s_{\alpha+\beta}} ~y_{h_0W^\prime W^\prime}.
\label{eq:ygauge}
\end{eqnarray}
We see that $s_{\alpha+\beta}$ is the most important parameter controlling these scaling factors and note that the coupling to the $W^\prime$ gauge boson is suppressed at $\mathcal{O}(v^2/F^2)$.

The heavier Higgs states, $A_0$ and $H_0$, may also decay to combinations of scalar and gauge states, such as $A_0 \rightarrow h Z^{(*)}$ and $H_0 \rightarrow A_0 Z^{(*)}, H^\pm W^{\mp(*)}$. These couplings do not have an equivalent expression in the SM, so we provide the explicit BLH coupling expressions below:
\begin{eqnarray} \label{g_SSV}
g_{A_0Zh_0} &=& i \frac{g}{2c_w} c_{\alpha+\beta} +\mathcal{O}(v^2/f^2) \cr\cr
g_{H_0ZA_0} &=&  -i \frac{g}{2c_w} s_{\alpha+\beta} +\mathcal{O}(v^2/f^2)  \cr\cr
g_{H_0H^\pm W^\mp} &=&  \mp \frac{g}{2} s_{\alpha+\beta} +\mathcal{O}(v^2/f^2)
\end{eqnarray}

\subsection{Fermion Sector}

\begin{figure*}[htbp]
   \begin{center}
	\includegraphics[width=0.6\textwidth]{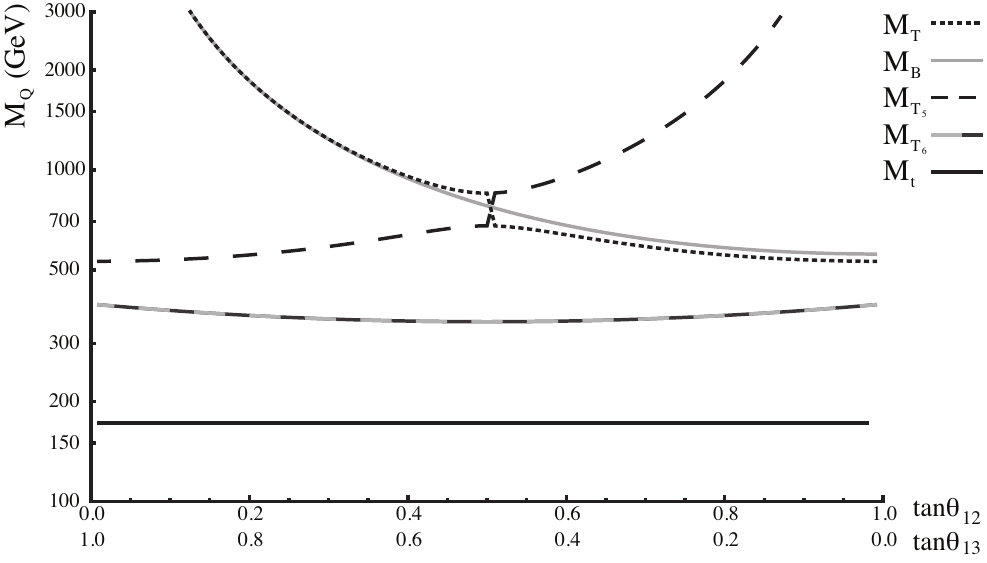}
   \end{center}
   \vspace{-4mm}
   \caption{Slice of fermion mass plots, depending on mixing angles $\tan\theta_{12}$ and $\tan\theta_{13}$, corresponding to ($f=1000$~GeV, $\tan\beta=3.35$). The point at which the heaviest state switches from the $T$ to $T_5$ is obvious.}
\label{fig:fmass}
\end{figure*}

The Yukawa terms for the light fermions in the BLH model Lagrangian take the form 
\begin{equation}
\mathcal{L}=y_u f Q^T \Sigma U^c +y_d f Q^T (-2 i T_R^2 \Sigma) D^c + h.c. ,
\end{equation}
where $T_R^2$ is the second component of the triplet of $SU(2)_R$ generators corresponding to the $SO(4)_R$ subgroup of $SO(6)$, and performs the charge conjugation of the Higgs fields for interactions with the down-type quarks. This identifies the BLH model as a Type I 2HDM, with light fermion masses given by
\begin{equation}
m_f^2 = y_f^2 v^2 \sin^2\beta \left(1-\frac{v^2}{3f^2}+...\right).
\end{equation}
and scaling factors of the form
\begin{eqnarray}
y_{h_0 f \bar{f}} &=& \frac{c_\alpha}{s_\beta}-\frac{2v^2}{3f^2} s_{\alpha+\beta} \cr\cr
y_{H_0 f \bar{f}} &=& -\frac{s_\alpha}{s_\beta}-\frac{2v^2}{3f^2} c_{\alpha+\beta} \cr\cr
y_{A_0 u \bar{u}} &=& \cot{\beta}\left(1+ \frac{2 v^2 }{3f^2 }\right)\cr\cr
y_{A_0 d \bar{d}} &=& -y_{A_0 u \bar{u}} ~.
\end{eqnarray}
The following expressions give the leading order contributions in terms of $s_{\alpha+\beta}$ and $t_\beta$:
\begin{eqnarray}
\frac{c_\alpha}{s_\beta} &=& s_{\alpha+\beta}+ c_{\alpha+\beta}/ t_{\beta}, \cr \cr
\frac{s_\alpha}{s_\beta} &=& -c_{\alpha+\beta}+ s_{\alpha+\beta}/ t_{\beta}.
\end{eqnarray}

The BLH model also includes a number of heavy, vector-like quarks that act to protect the Higgs boson mass from developing quadratic divergences from the top quark loops. The Yukawa interactions for the heavy fermions, which includes the top quark, take the form given in Eq. 46 of \cite{Schmaltz:2010ac}:
\begin{equation}
\mathcal{L}=y_1 f Q^T S \Sigma S U^c + y_2 f Q_a^{\prime T} \Sigma U^c + y_3 f Q^T \Sigma U_5^{\prime c}+ h.c. .
\end{equation}
These couplings can be parameterized in terms of $y_1$ and two mixing angles, defined such that $y_2 = y_1/\tan\theta_{12}$ and $y_3 = y_1/\tan\theta_{13}$.  The value of $y_1$ is then fixed through the measured top quark mass.

As we showed in \cite{Godfrey:2012tf}, analytic methods for determining the mass eigenstates for the heavy quarks fail in the region where $| y_2 - y_3 | \approx 0$, due to the degeneracy between the $T$ and $T_5$ states. This can clearly be seen by observing a slice of the $(\tan\theta_{12},\tan\theta_{13})$ parameter space, where $\tan\theta_{13} = 1- \tan\theta_{12}$, as in Fig. \ref{fig:fmass}. As a result, we use numerical diagonalization to determine the mass eigenstates of the heavy quarks, and thus their coupling to the $h_0$, $H_0$ and $A_0$ states, for all values of $\tan\theta_{12}$ and $\tan\theta_{13}$.  Therefore, we do not provide analytic solutions for the relevant scaling factors.

However, we can provide some insight into their contributions to the Higgs production and decay. Since the vector-like quarks ($T, B, T_b^{2/3}, T_b^{5/3}, T_5, T_6$) obtain contributions to their masses proportional to $f$, the lowest order contribution to their scaling factor is $\mathcal{O}(v/f)$, and their contribution to the loops in the $S_0 gg$ and $S_0 \gamma \gamma$ effective vertices drop off rapidly with increasing $f$. In particular, the scaling factors for the $B$ and $T_b^{5/3}$ are suppressed at $\mathcal{O}(v^3/f^3)$, making these two states effectively decouple from the process. The scaling factors for the $T_5$ and $T$ are the largest for the set, but with values typically smaller than 0.02, while for the $T_6$ and $T_b^{2/3}$ the scaling factor is suppressed by an additional factor of about 10. 

Since the top quark is generated from the heavy quark Yukawa terms, it gets a mass proportional to $v$, and thus the dependence of the scaling factor for the top quark on $\beta$ and $\alpha$ behaves at lowest order like those of the light up-type quarks. This dependence on $\beta$ and $\alpha$, characteristic of a Type I 2HDM, results in the most significant deviations from the SM in the fermion contributions to the loop interactions.

\subsection{Parameter Survey}

We use the following fixed values in our calculations \cite{PDG2012}: 
\begin{eqnarray}
\alpha_{EM} &=& 1/127.9\cr
G_F &=& 0.0000116637~\mathrm{GeV^{-2}}\cr
\widehat{m}_Z &=& 91.1876~\mathrm{GeV}\cr
m_t &=& 172.5~\mathrm{GeV}\cr
m_b &=& 4.16~\mathrm{GeV}\cr
m_\tau &=& 1.77684~\mathrm{GeV}\cr
m_c &=& 1.28~\mathrm{GeV}\cr
m_4 &=& 30~\mathrm{GeV}\cr
m_5 &=& 30~\mathrm{GeV}\cr
m_6 &=& 30~\mathrm{GeV}
\end{eqnarray}
Of note, the parameters $m_{4,5,6}$ are explicit symmetry breaking mass terms in the full BLH model scalar potential \cite{Schmaltz:2010ac} that are small and provide a negligible contribution to the interactions examined. As discussed in \cite{Schmaltz:2010ac}, these parameters are introduced to break all the axial symmetries in the Higgs potential, giving positive masses to all scalars.  In particular, the $\eta_0$ state receives a mass equal to $m_4 \sim 10$~GeV.  Since this state couples to top quarks and would decay predominantly to $b$ quarks, it may be visible in $t\bar{t}b\bar{b}$ final states.  The value of $m_4$
 would strongly affect the rate for this process, but this is not relevant for the signal strengths considered here.

To be thorough, we randomize all remaining parameters in the BLH model over the ranges:
\begin{eqnarray}
m_{h_0} &\in& (124,126)~\mathrm{GeV}\cr
m_{A_0} &\in& (m_{h_0},~\mathrm{Max}(m_{A_0}))\;\;\mathrm{(see\,Eq.\,\ref{eq:mA0max})}\cr
f &\in& (700,3000)~\mathrm{GeV}\cr
F &\in& (\mathrm{Min}(F),\mathrm{Min}(F)+ 4000~\mathrm{GeV})\cr
\tan\beta &\in& (1,\mathrm{Max}(\tan\beta))\;\;\mathrm{(see\,Eq.\,\ref{eq:tBmax})}\cr
\tan\theta_g &\in& (0,5)\cr
\tan\theta_{12} &\in& (0,5)\cr
\tan\theta_{13} &\in& (0,5)\cr
K_\sigma &\in& (1,\mathrm{Max}(K_\sigma))\;\; \mathrm{(see\,Eq.\,\ref{eq:sigmamax})}\cr
\kappa_G &\in& (0,5)\cr
\kappa_Y &\in& (0,5)\cr
\kappa_S &\in& (0,5)
\end{eqnarray}
where
 \begin{equation}~\mathrm{Max}(m_{A_0}) =  \left\{ \begin{array}{l l} 700~\mathrm{GeV} & \textrm{(general scenario)} \\
                            128~\mathrm{GeV}  & \textrm{(near-degenerate scenario).}
                \end{array} \right.  
                \label{eq:mA0max}
\end{equation}
We determine Min($F$) as follows. Electroweak precision observables (EWPO) place 
constraints on the mass of the heavy gauge bosons, as a function of the heavy gauge boson 
mixing angle, $\sin\theta_g$, as shown in \cite{Schmaltz:2010ac} for a light Higgs boson 
mass of 125 GeV. This lower limit on the $W^\prime$ mass in turn determines Min($F$) for 
each value of $\sin\theta_g$ via the third line in Equation \ref{gauge_masses}. So the parameter sets we 
generate satisfy this EWPO constraint.

Since the mass of the observed resonance is not precisely known, and especially since the $ZZ^*$ and $\gamma\gamma$ mass peak values in the ATLAS measurements are distinct, we allow the mass of the light Higgs boson to vary over a small range. All other values are calculated from this base set, as described in the preceding sections of this paper.

Additionally, we separate our analysis into two scenarios - a ``general case", where $m_{A_0}$ varies up to 700 GeV, and a ``near-degenerate" case, where $m_{A_0}$ can take a maximum value of $128$~GeV.

\section{Results \label{sec:results}}

Both CMS and ATLAS have now published updated signal strengths for each given production mode and final state using the 7 TeV and 8 TeV data. We have calculated the expected value for these signal strength ratios in the BLH model for 60k parameter sets for each of the general and the near-degenerate cases. Each set corresponds to a point on the plots which follow. Specifically, we calculate:
\begin{eqnarray*}
\mu^{BLH}_{XX} = \displaystyle\frac{\mathcal{L}_7 \displaystyle\sum_i \sigma_{i,7}^{BLH}BR_{XX}^{BLH} + \mathcal{L}_8 \displaystyle\sum_i \sigma_{i,8}^{BLH} BR_{XX}^{BLH}}{\left(\mathcal{L}_7 \displaystyle\sum_i \sigma_{i,7}^{SM}BR_{XX}^{SM} + \mathcal{L}_8 \displaystyle\sum_i \sigma_{i,8}^{SM} BR_{XX}^{SM}\right)\bigg|_{m_h^{exp}}}
\end{eqnarray*}
where
\begin{eqnarray}
BR_{XX} &=& \displaystyle\frac{\Gamma(S_0 \rightarrow XX)}{\displaystyle\sum_Y \Gamma(S_0 \rightarrow YY)}.
\label{eq:mu}
\end{eqnarray}
The sum over the index $i$ accounts for all contributing production modes, including the $A_0$ and $H_0$ mediated production. Since the signal strengths from CMS and ATLAS are calculated based on the experimentally determined best fit mass, our $\mu$ values are normalized to the expected results for a SM Higgs boson with the mass given in the experimental study, and are weighted by the 7 TeV and 8 TeV integrated luminosities, $\mathcal{L}_7$ and $\mathcal{L}_8$ respectively. Summaries of the published results from CMS and ATLAS can be found in Tables \ref{tab:CMSresults} and \ref{tab:ATLASresults}, respectively.

While we include contributions from the $A_0$ and $H_0$ in our determination of the signal strength ratios, some constraints currently exist on heavy scalars. In particular, both CMS~\cite{Chatrchyan:2013yoa} and ATLAS~\cite{TheATLAScollaboration:2013zha} have placed 95\% C.L. constraints on the decays of a heavy Higgs boson to $W^+W^-$, and CMS provides 95\% C.L. exclusions on $\sigma/\sigma_{SM}$ for the $H \rightarrow \gamma\gamma$ final state for masses up to 150 GeV~\cite{CMS:ril}. We have incorporated these results and show parameter points that violate these constraints in light red/pink to distinguish them from the unconstrained results.

We present our results using two different $\chi^2$ measures in Figures \ref{fig:rdeg:muyyvsmuww} through \ref{fig:norm:flvsmT6}. In order to show the BLH parameter sets that produce results in better agreement with the measured data than the SM, we calculate $\chi^2_{BLH} - \chi^2_{SM}$, where
\begin{equation}
\chi^2_{BLH/SM} = \displaystyle\sum_i (\chi^2_{BLH/SM})_i= \sum_i \frac{(\mu_i^{BLH/SM}-\mu_i^{obs})^2}{(\delta\mu_i^{obs})^2},
\end{equation}
where the sum over the index $i$ includes all channels listed in Tables \ref{tab:CMSresults} and \ref{tab:ATLASresults} and $\mu_i^{SM} \equiv 1$. This measure is employed in Figures \ref{fig:deg:tBvssapb}, \ref{fig:norm:tBvssapb}-\ref{fig:norm:muwwvsmuzz}, and \ref{fig:norm:flvsmT6}. For Figures \ref{fig:rdeg:muyyvsmuww}-\ref{fig:rdeg:tBvssapb}, \ref{fig:rdeg:muyyvsmutata}, and \ref{fig:rnorm:muyyvsmuww}-\ref{fig:rnorm:hmuyy1vsAmuyy1}, we calculate an alternate measure $\Delta\chi^2$, defined as:
\begin{equation}
\Delta\chi^2 = \displaystyle\sum_{i = \gamma\gamma, WW^*}(\chi^2_{BLH})_i - \chi^2_{min},
\end{equation}
where the sum in this case is only over the $\gamma\gamma$ and $WW^*$ channels from Tables \ref{tab:CMSresults} and \ref{tab:ATLASresults} and
\begin{equation} \label{chisq_min}
\chi^2_{min} = \textrm{Min}\left(\displaystyle\sum_{i = \gamma\gamma,WW^*}(\chi^2_{BLH})_i \right) ~.
\end{equation}
This measure focuses only on the most precisely measured signal strength values in order to identify regions of the BLH parameter space which favour either the distinctive CMS diphoton and $WW^*$ results or those of ATLAS. The values of $ \chi^2_{min}$ in the near-degenerate and general scenarios are listed in Table \ref{tab:chisq}. In all figures, both the ATLAS (right panel) and CMS (left panel) results are included, in order to show the differences between the ATLAS and CMS preferred regions.

\begin{table*}[t]
\begin{center}
\caption{Summary of published Higgs boson results from CMS. Note: The quoted $\gamma\gamma$ result uses the multi-variate analysis value, rather than the cut based analysis value.}
\label{tab:CMSresults}
\begin{tabular}{ | c c c c c c | }
\hline
Signal & Production & $\mathcal{L}_7$ (fb$^{-1}$) & $\mathcal{L}_8$ (fb$^{-1}$) & $\hat{\mu}$ & $m_{h}$ (GeV) \\\hline
$\gamma\gamma$\cite{CMS:ril} & Inclusive & 5.1 & 19.6 & $0.78_{-0.26}^{+0.28}$ & 125\\
$ZZ^*$\cite{CMS:xwa} & Inclusive & 5.1 & 19.6 & $0.91_{-0.24}^{+0.30}$ & 125.8\\
$ZZ^*$\cite{CMS:xwa} & VBF & 5.1 & 19.6 & $1.22_{-0.57}^{+0.84}$ & 125.8\\
$WW^*$\cite{CMS:bxa} & ggF & 4.9 & 19.5 & $0.76\pm0.21$ & 125\\
$b\bar{b}$\cite{CMS:cya} & VH & 5.0 & 12.1 & $1.3_{-0.6}^{+0.7}$ & 125\\
$\tau^+\tau^-$\cite{CMS:utj} & Inclusive & 4.9 & 19.4 & $1.1\pm0.4$ & 125\\
$\tau^+\tau^-$\cite{CMS:utj} & VBF & 4.9 & 19.4 & $1.4\pm0.6$ & 125\\\hline
\end{tabular}

\end{center}
\end{table*}

\begin{table*}[t]
\begin{center}
\caption{Summary of published Higgs boson results from ATLAS.}
\label{tab:ATLASresults}
\begin{tabular}{ | c c c c c c | }
\hline
Signal & Production & $\mathcal{L}_7$ (fb$^{-1}$) & $\mathcal{L}_8$ (fb$^{-1}$) & $\hat{\mu}$ & $m_{h}$ (GeV) \\\hline
$\gamma\gamma$\cite{ATLAS:2013oma} & Inclusive & 4.8 & 20.7 & $1.65_{-0.30}^{+0.34}$ & 126.8\\
$ZZ^*$\cite{ATLAS:2013nma} & Inclusive & 4.6 & 20.7 & $1.7_{-0.4}^{+0.5}$ & 124.3\\
$WW^*$\cite{ATLAS:2013wla} & ggF+VBF & 4.6 & 20.7 & $1.01\pm0.31$ & 125\\
$WW^*$\cite{ATLAS:2013wla} & VBF & 4.6 & 20.7 & $1.66\pm0.79$ & 125\\
$WW^*$\cite{ATLAS:2013wla} & ggF & 4.6 & 20.7 & $0.82\pm0.36$ & 125\\
$b\bar{b}$\cite{ATLAS:2012wma} & VH & 4.7 & 13.0 & $-0.4\pm1.0$ & 125\\
$\tau^+\tau^-$\cite{ATLAS:2012dsy} & Inclusive & 4.6 & 13.0 & $0.7\pm0.7$ & 125\\\hline

\end{tabular}
\end{center}
\end{table*}

\begin{table}
\begin{center}
\caption{Values of $\chi^2_{min}$, as defined in Eq. \ref{chisq_min}, in the general and near-degenerate cases.
These calculated values include only the $\gamma\gamma$ and $WW^*$ channels from Tables \ref{tab:CMSresults} and \ref{tab:ATLASresults}. }
\label{tab:chisq}
\begin{tabular}{| l c | l c |}
\multicolumn{2}{c}{Near-degenerate} & \multicolumn{2}{c}{General} \\\hline
Experiment & $\chi^2_{min}$ & Experiment & $\chi^2_{min}$ \\\hline
CMS 	& 0.001 & 	CMS 	& 0.007 \\
ATLAS 	& 0.08 & 	ATLAS 	& 0.11 \\\hline
\end{tabular}
\end{center}
\end{table}

\subsection{Near-Degenerate Scenario}

\begin{figure*}[tp]
   \begin{center}
	\includegraphics[width=\textwidth]{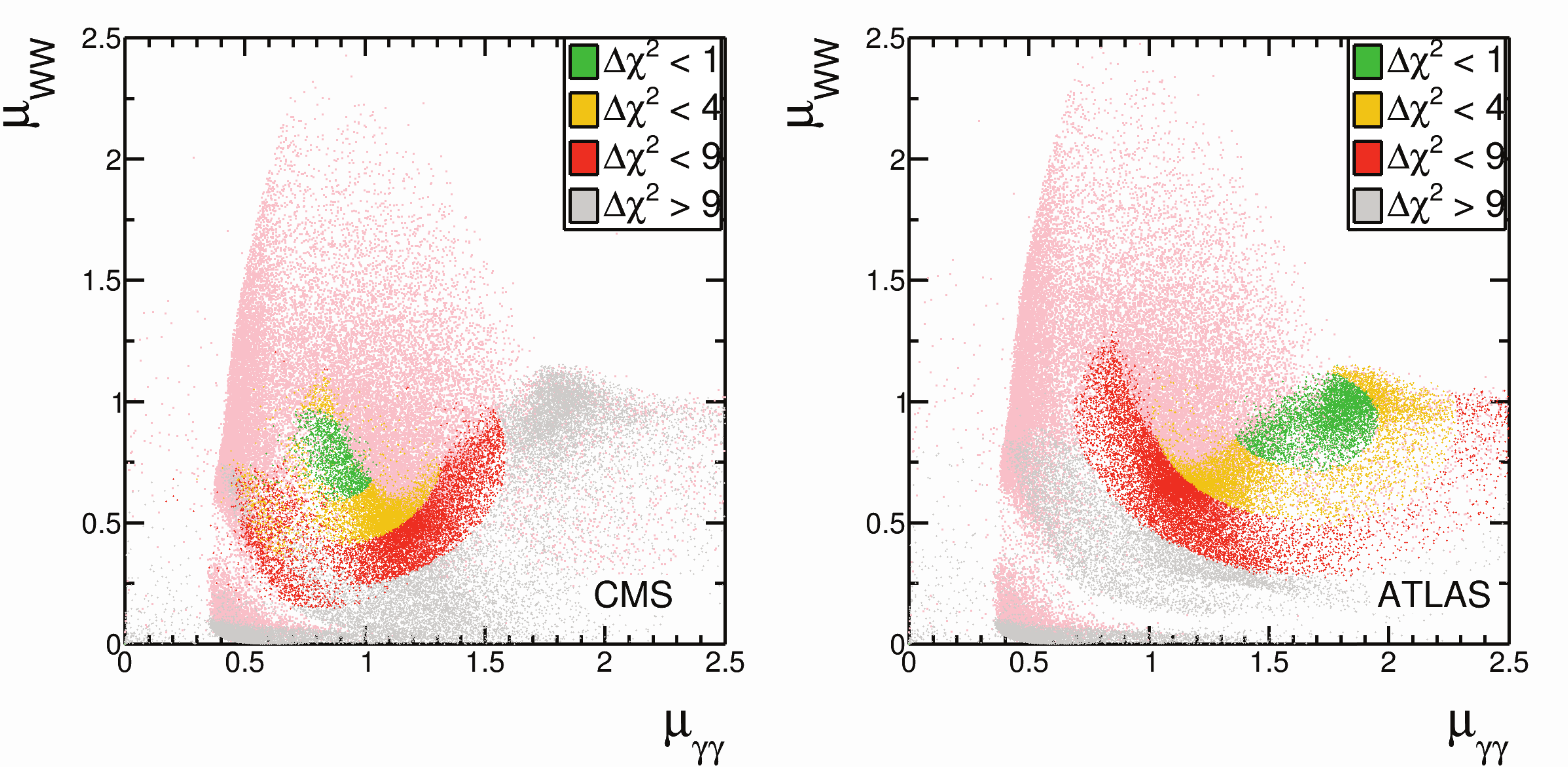}
   \end{center}
   \caption{Near-Degenerate Scenario: $WW^*$ versus diphoton signal strength ratios, assuming a reduced $\Delta\chi^2$ calculation (including only $\mu_{\gamma\gamma}$ and $\mu_{WW}$ measurements) to focus on the disparity between the CMS and ATLAS diphoton signal strength ratios. Parameter points in pink indicate an exclusion at 95\% C.L. due to high mass resonance.}
\label{fig:rdeg:muyyvsmuww}
\end{figure*}

\begin{figure*}[tp]
   \begin{center}
	\includegraphics[width=\textwidth]{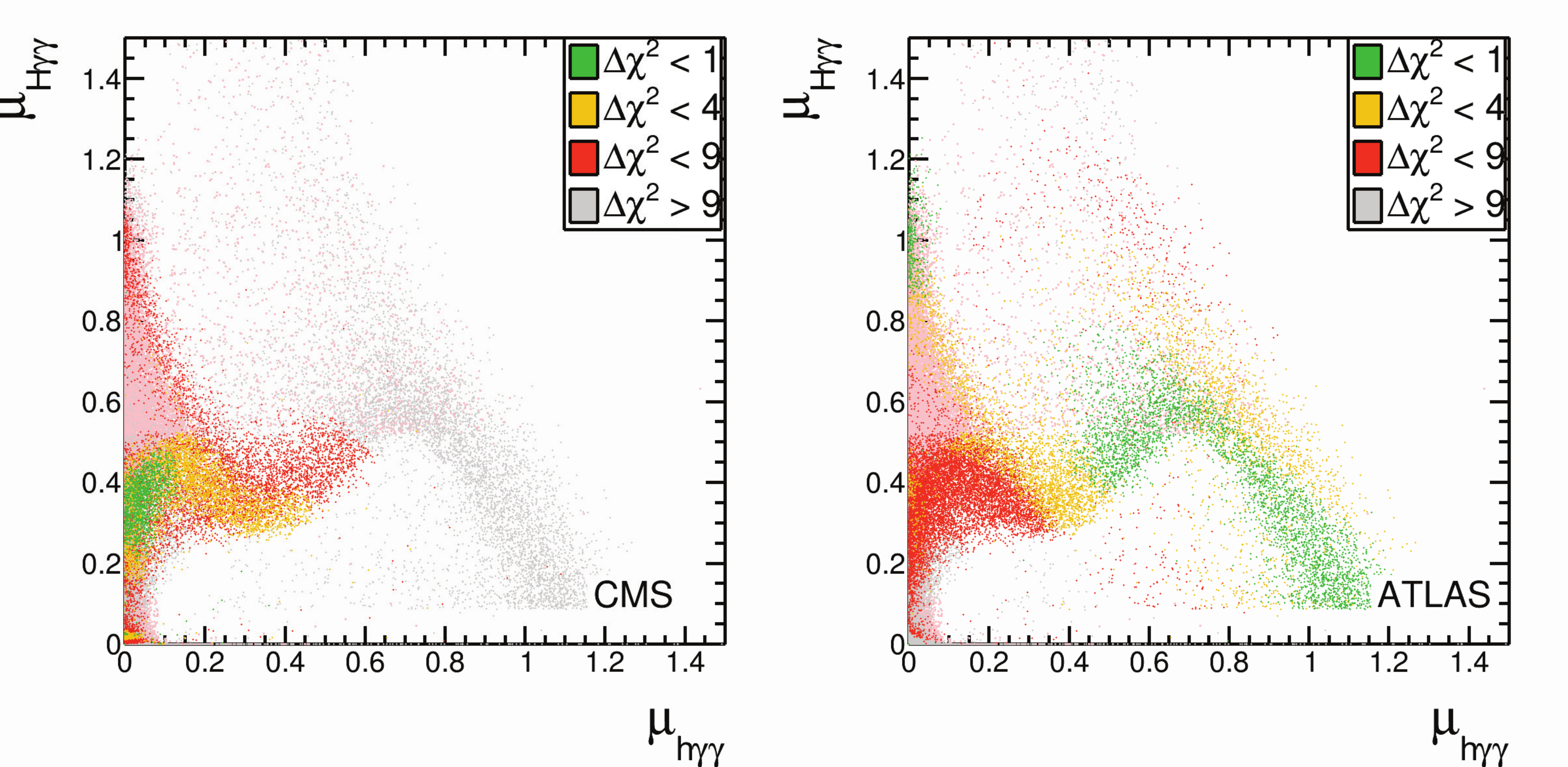}
   \end{center}
   \caption{Near-Degenerate Scenario: Heavy versus light Higgs boson contributions to the diphoton signal strength ratio, assuming a reduced $\Delta\chi^2$ calculation (including only $\mu_{\gamma\gamma}$ and $\mu_{WW}$ measurements) to focus on the disparity between the CMS and ATLAS diphoton signal strength ratios. Parameter points in pink indicate an exclusion at 95\% C.L. due to high mass resonance.}
\label{fig:rdeg:hmuyyvsHmuww}
\end{figure*}

\begin{figure*}[tp]
   \begin{center}
	\includegraphics[width=\textwidth]{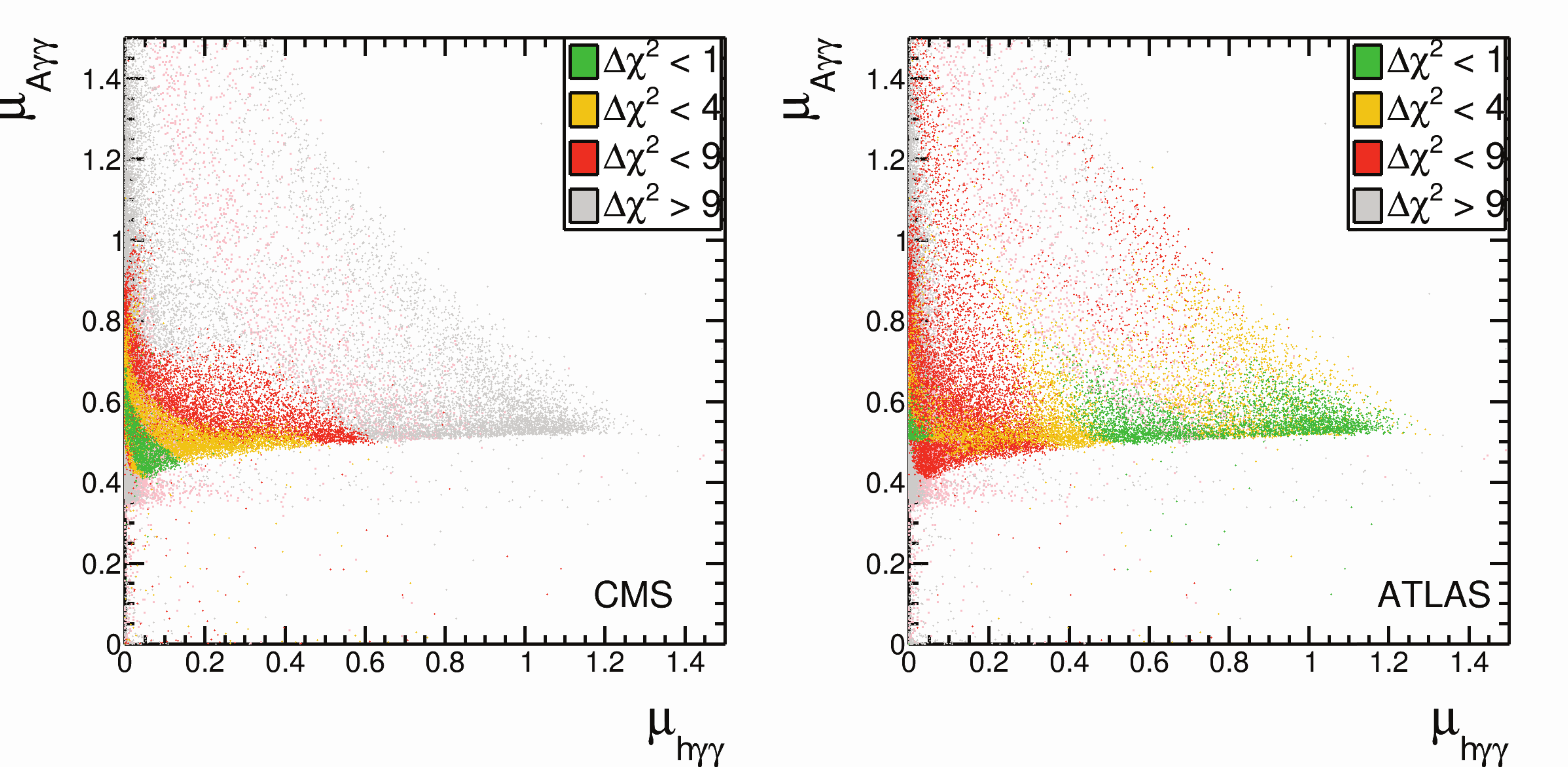}
   \end{center}
   \caption{Near-Degenerate Scenario: CP-odd scalar versus light Higgs boson contributions to the diphoton signal strength ratio, assuming a reduced $\Delta\chi^2$ calculation (including only $\mu_{\gamma\gamma}$ and $\mu_{WW}$ measurements) to focus on the disparity between the CMS and ATLAS diphoton signal strength ratios. Parameter points in pink indicate an exclusion at 95\% C.L. due to high mass resonance.}
\label{fig:rdeg:hmuyyvsAmuww}
\end{figure*}

\begin{figure*}[tp]
   \begin{center}
	\includegraphics[width=\textwidth]{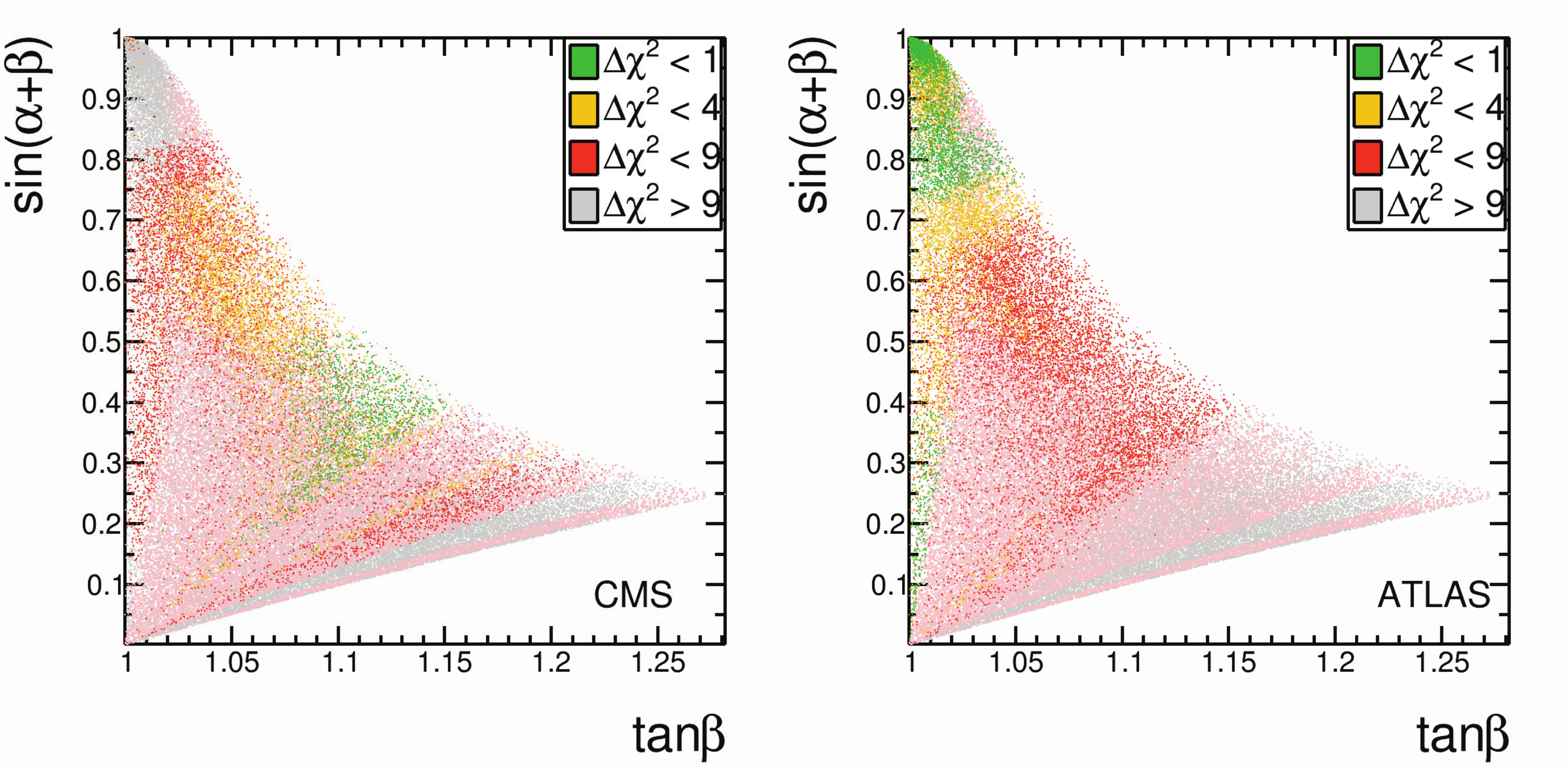}
   \end{center}
   \caption{Near-Degenerate Scenario: Comparison of the $\sin(\alpha+\beta)$ and $\tan\beta$ parameters, assuming a reduced $\Delta\chi^2$ calculation (including only $\mu_{\gamma\gamma}$ and $\mu_{WW}$ measurements) to focus on the disparity between the CMS and ATLAS diphoton signal strength ratios. Parameter points in pink indicate an exclusion at 95\% C.L. due to high mass resonance.}
\label{fig:rdeg:tBvssapb}
\end{figure*}

\begin{figure*}[tp]
   \begin{center}
	\includegraphics[width=\textwidth]{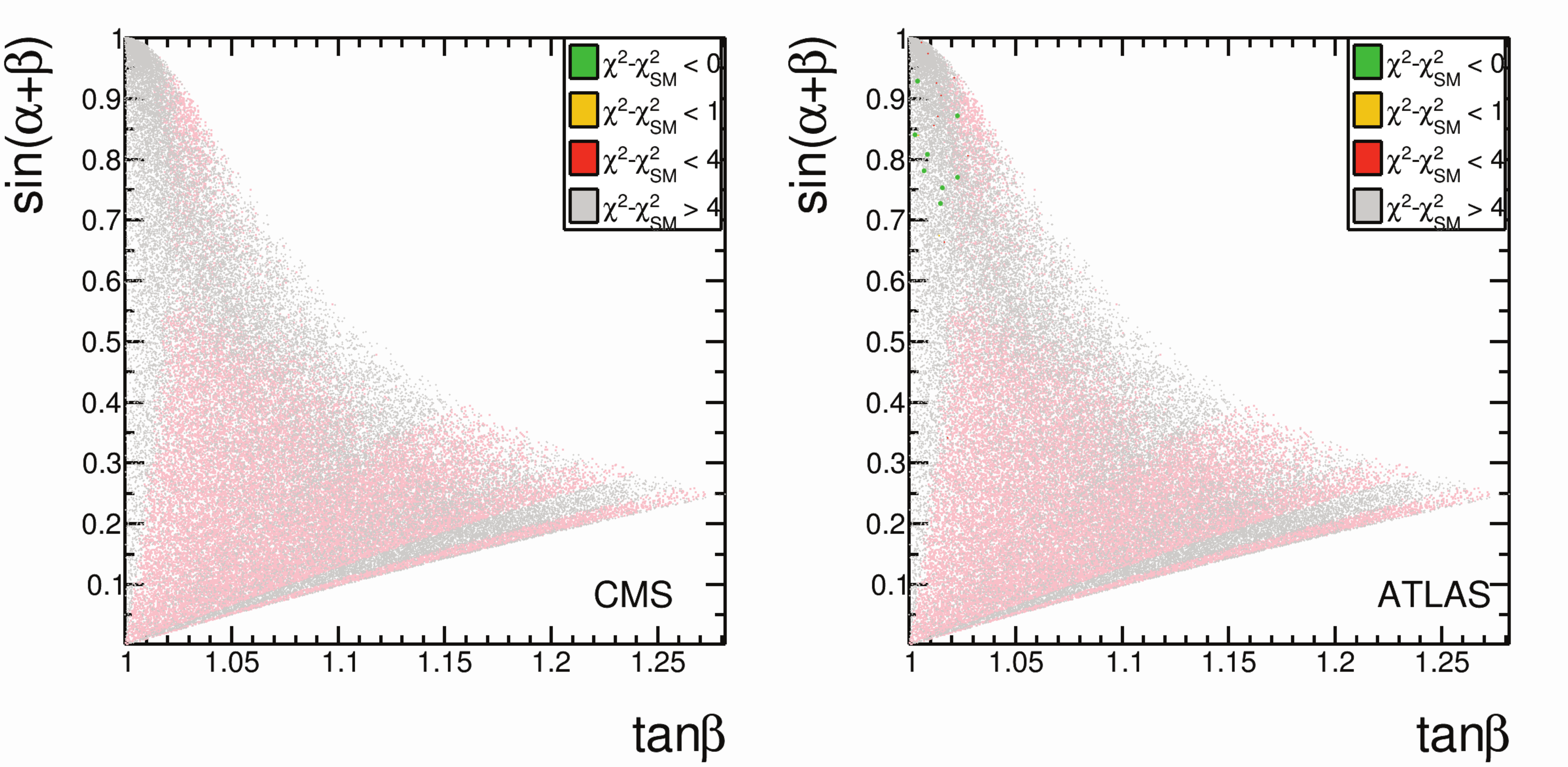}
   \end{center}
   \caption{Near-Degenerate Scenario: Comparison of the $\sin(\alpha+\beta)$ and $\tan\beta$ parameters, using a $\chi^2 - \chi^2_{SM}$ measure to compare the BLH model predictions to the SM predictions, including the full set of measured signal strength ratios. Parameter points in pink indicate an exclusion at 95\% C.L. due to high mass resonance.}
\label{fig:deg:tBvssapb}
\end{figure*}

\begin{figure*}[tp]
   \begin{center}
	\includegraphics[width=\textwidth]{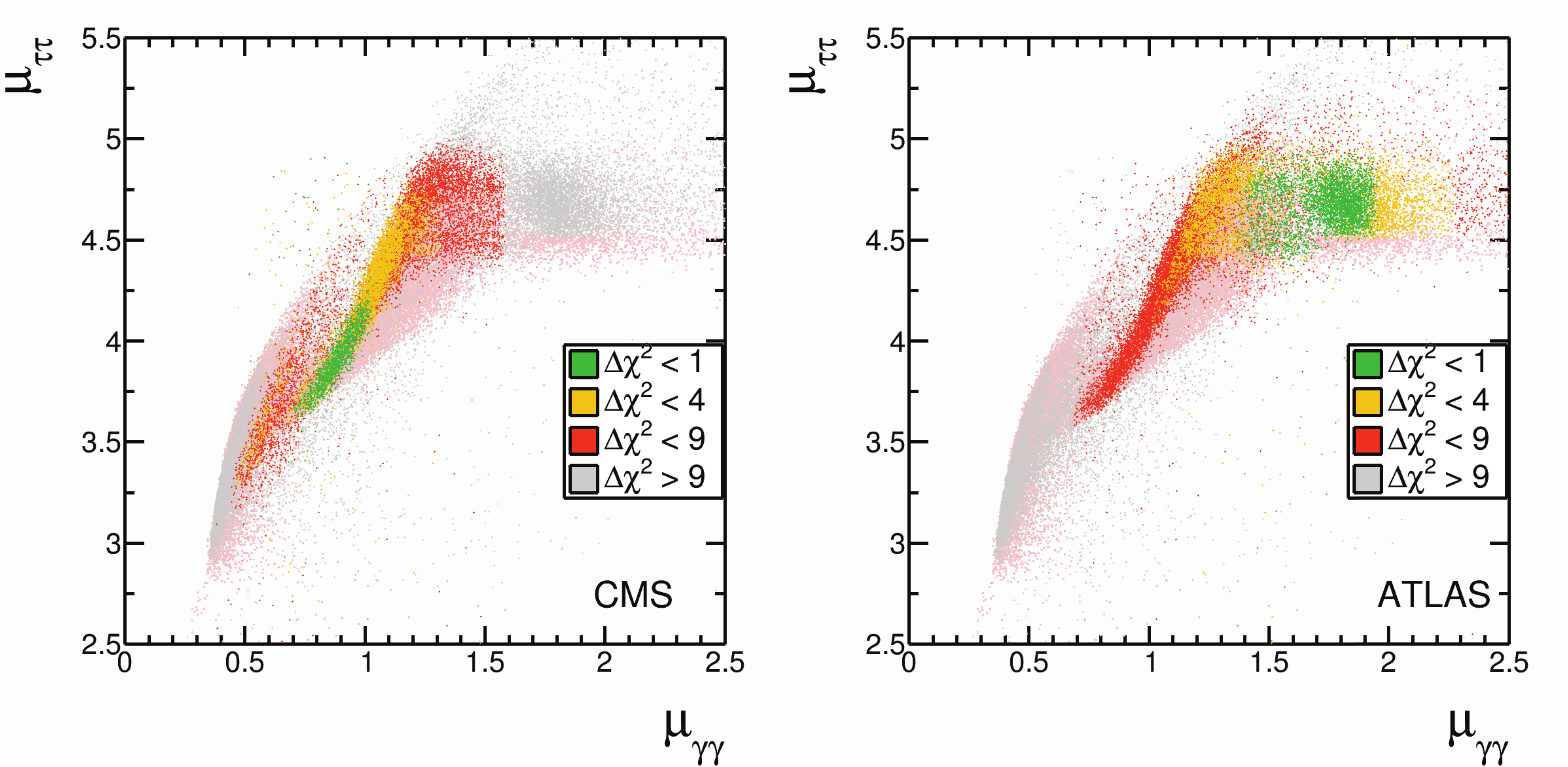}
   \end{center}
   \caption{Near-Degenerate Scenario: Comparison of the $\tau^+\tau^-$ signal strength ratio to the diphoton ratio, assuming a reduced $\Delta\chi^2$ calculation (including only $\mu_{\gamma\gamma}$ and $\mu_{WW}$ measurements) to focus on the disparity between the CMS and ATLAS diphoton signal strength ratios. Parameter points in pink indicate an exclusion at 95\% C.L. due to high mass resonance.}
\label{fig:rdeg:muyyvsmutata}
\end{figure*}

We define a near-degenerate scenario such that the mass of the $A_0$ is constrained to be 
within a few GeV of the mass of the lighter scalar Higgs state ($h_0$), which allows it to 
contribute to any signal strength that involves fermionic couplings in both the production 
mode (ggF, ttH) and the decay mode ($b\bar{b}$, $\tau^+\tau^-$, $\gamma\gamma$ and 
$Z\gamma$). Depending on the other parameters, the mass of the heavier scalar Higgs state 
($H_0$) can also be nearly degenerate with the $h_0$, and thus also contribute significantly 
to the measured signal strengths, including $WW^*$ and $ZZ^*$. Ultimately, this scenario 
can reproduce an enhanced diphoton signal strength, as measured by ATLAS, due to 
multiple scalar states contributing. However, at the same time, the $\tau^+\tau^-$ mode is 
also enhanced, beyond what is experimentally observed. So the scenario is largely ruled out. 
However, it remains useful as a tool to study the contribution of the different scalar states to 
the various processes.

Since several scalar fields may be contributing significantly to the diphoton production rate, it 
is useful to consider the individual contribution to $\mu_{\gamma\gamma}$ from each of the 
$h_0$, $A_0$ and $H_0$. For this purpose, we define
\begin{eqnarray}
\mu_{\gamma\gamma} &=& \mu_{h\gamma\gamma} + \mu_{A\gamma\gamma} + 
\mu_{H\gamma\gamma},
\end{eqnarray}
where $\mu_{S\gamma\gamma}$ includes only the cross section from the production and 
decay of the stated scalar, $S$.  

We first focus on the issue of the CMS results versus those of ATLAS. Figure \ref{fig:rdeg:muyyvsmuww} shows that both the more SM-like $\mu_{\gamma\gamma}$ from CMS and the excess $\mu_{\gamma\gamma}$ from ATLAS can be accommodated in the near-degenerate scenario. Together, Figures \ref{fig:rdeg:hmuyyvsHmuww}, \ref{fig:rdeg:hmuyyvsAmuww} and \ref{fig:rdeg:tBvssapb} provide insight into the reproduction of the data of each experiment within the BLH parameter space, in terms of the particular scalar states that are being produced, and in terms of the parameters $\sin(\alpha+\beta)$ and $\tan\beta$. 
Those parameter sets that yield results consistent with the CMS data involve a strongly suppressed $h_0$ contribution to the diphoton signal strength ratio, $\mu_{h\gamma\gamma}$, as seen in the left panels of Figures \ref{fig:rdeg:hmuyyvsHmuww} and \ref{fig:rdeg:hmuyyvsAmuww}. The reduced $h_0$ contribution results from a suppression of the coupling between the $h_0$ and the $W$ boson that occurs for small to moderate values of $\sin(\alpha+\beta)$ (see equation \ref{eq:ygauge} for $\alpha$ and $\beta$ dependence), as seen in the left panel of Figure \ref{fig:rdeg:tBvssapb}. The $h_0$ coupling to the top quark is slightly enhanced for this range of $\sin(\alpha+\beta)$ and $\tan\beta$ values, which results in destructive interference with the gauge boson contribution to the diphoton effective coupling. The diphoton rate as measured by CMS, which is near that of the SM, is instead understood as predominantly a result of contributions from both the $A_0$ and $H_0$ decays. The $WW^*$ signal strength ratio, as measured by CMS, is produced from a combination of the decays of the $h_0$ and $H_0$.

The right panels of Figures \ref{fig:rdeg:hmuyyvsHmuww}, \ref{fig:rdeg:hmuyyvsAmuww} and \ref{fig:rdeg:tBvssapb} give equivalent information for the reproduction of the ATLAS data. Here it is primarily BLH parameter sets with large $\sin(\alpha+\beta)$ and small $\tan\beta$ that are favoured by the data. This range of the parameters yield a diphoton signal ratio approximately 40\% due to production and decay of the CP-odd state, $A_0$, with an admixture of decays of the $H_0$ and $h_0$ providing the rest of the contribution. The $WW^*$ production arises primarily from the $h_0$, with a smaller contribution from the $H_0$, which is proportional to $\cos(\alpha+\beta)$.

The results presented in Figures \ref{fig:rdeg:muyyvsmuww} through \ref{fig:rdeg:tBvssapb} only include the diphoton and $WW^*$ signal rates in the $\chi^2$ measure. The picture is drastically different when accounting for the full data set, as shown in Figure \ref{fig:deg:tBvssapb}. Comparing the full $\chi^2_{BLH}$ to that of $\chi^2_{SM}$, it is clear that there are almost no parameter values in which the near-degenerate BLH model is a better fit to the data than is the SM, nor is it even close. This is due entirely to the $\mu_{\tau\tau}$ signal strength ratio, as shown in Figure \ref{fig:rdeg:muyyvsmutata}. When the CP-odd scalar field is nearly degenerate with the light Higgs boson, it contributes a significant amount to the production of $\tau^+\tau^-$ pairs. For most parameter regions producing a signal consistent with the diphoton rates observed by either CMS or ATLAS, the $\tau^+\tau^-$ signal strength ratio would need to be three to five times larger than the SM rate, and several sigma larger than the respective measured values.

These calculations do not include the effect of interference. There are no interference effects between the CP-even ($h_0$ and $H_0$) states and the CP-odd ($A_0$) state due to CP invariance. In the case where the heavier CP-even state, $H_0$, is nearly degenerate with the $h_0$, the mass difference between the two is typically large enough compared to the corresponding boson widths that the interference effects can be neglected. This was the approach in \cite{Ferreira:2012nv} wherein the authors point out that the experimental mass resolutions are significantly larger than the Higgs widths and, so, the assumption does not significantly constrain their analysis.  Since we are democratic in our scan, we do include some parameter points in which the mass separation is small enough such that interference effects should be included, however we find that the value of $\mu_{A\tau\tau}$, the contribution of the CP-odd scalar to the di-tau signal strength ratio, increases for decreasing $m_{H_0}-m_{h_0}$. The region with small $m_{H_0}-m_{h_0}$ is thus ruled out due to considerations of the CP-odd scalar contributions alone, suggesting that the inclusion of interference effects will not change our conclusions.

\subsection{General Scenario} \label{sec:general_scenario}

\begin{figure*}[tp]
\begin{center}
\includegraphics[width=\textwidth]{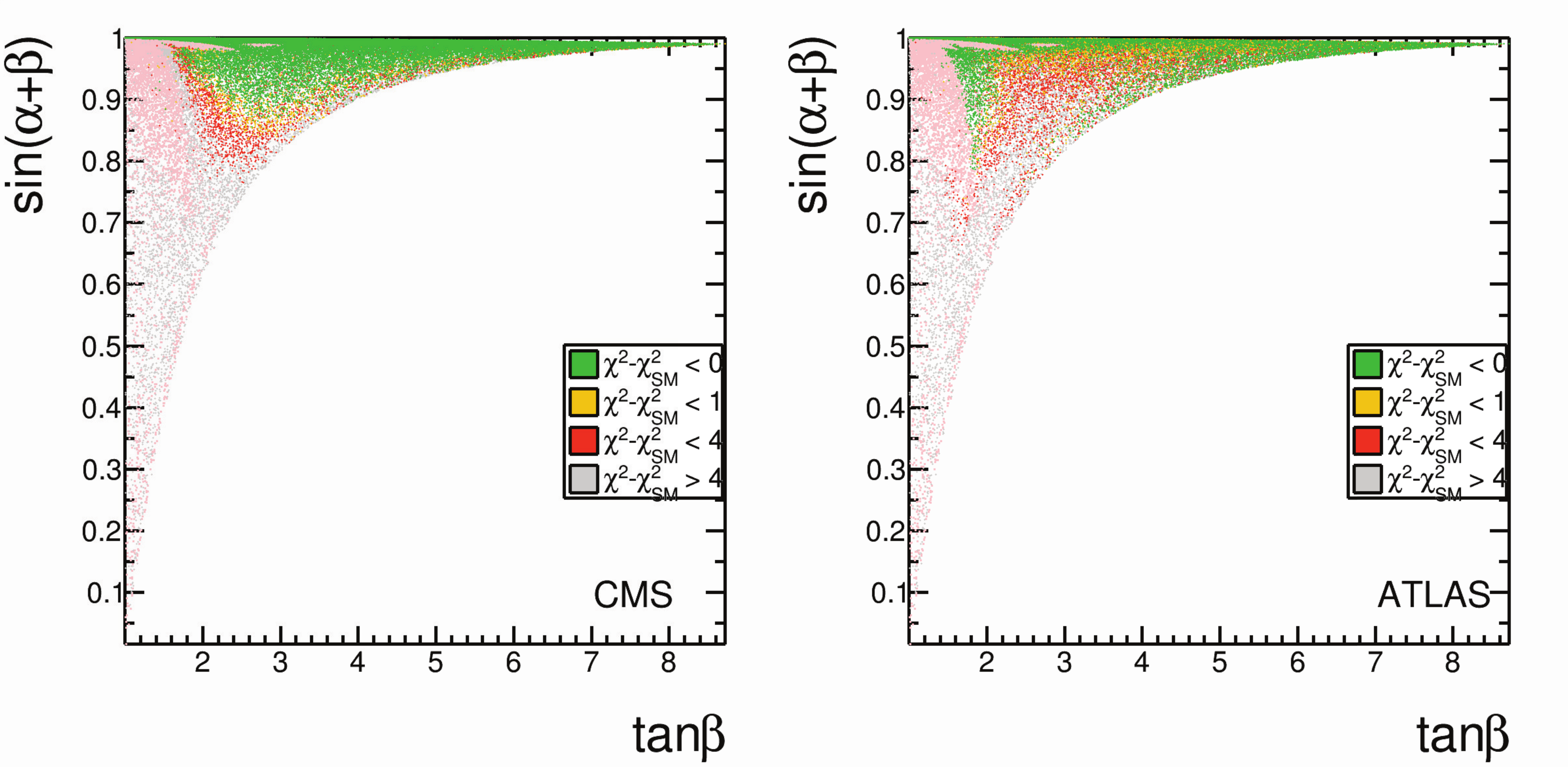}
\end{center}
\caption{General Scenario: Comparison of the $\sin(\alpha+\beta)$ and $\tan\beta$ parameters, using a $\chi^2 - \chi^2_{SM}$ measure to compare the BLH model predictions to the SM predictions, including the full set of measured signal strength ratios. Parameter points in pink indicate an exclusion at 95\% C.L. due to high mass resonance.}
\label{fig:norm:tBvssapb}
\end{figure*}

\begin{figure*}[tp]
\begin{center}
\includegraphics[width=\textwidth]{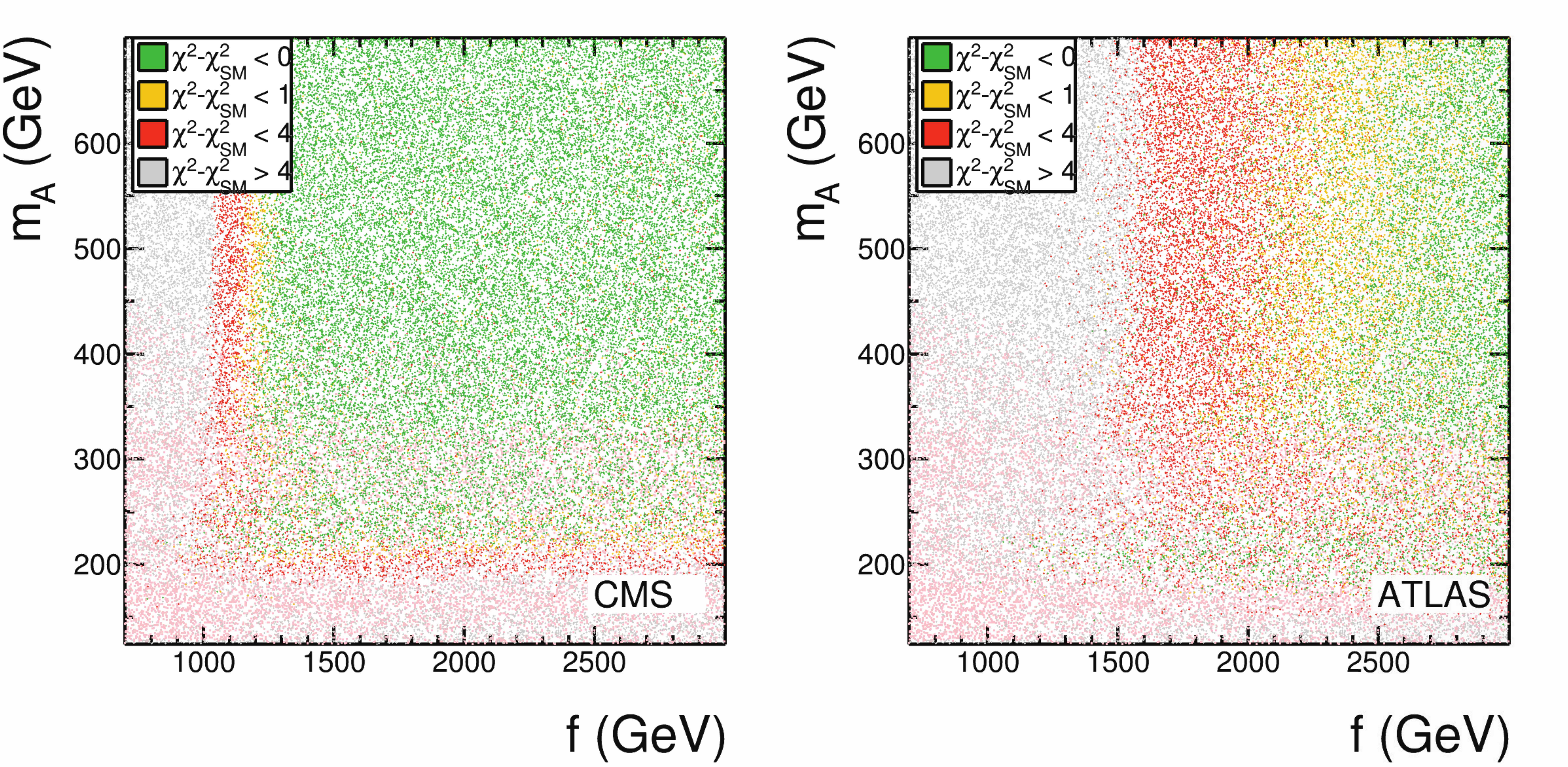}
\end{center}
\caption{General Scenario: Comparison of the CP-odd scalar mass, $m_{A_0}$, and $f$ parameters, using a $\chi^2 - \chi^2_{SM}$ measure to compare the BLH model predictions to the SM predictions, including the full set of measured signal strength ratios. Parameter points in pink indicate an exclusion at 95\% C.L. due to high mass resonance.
}
\label{fig:norm:flvsma0}
\end{figure*}

\begin{figure*}[tp]
\begin{center}
\includegraphics[width=\textwidth]{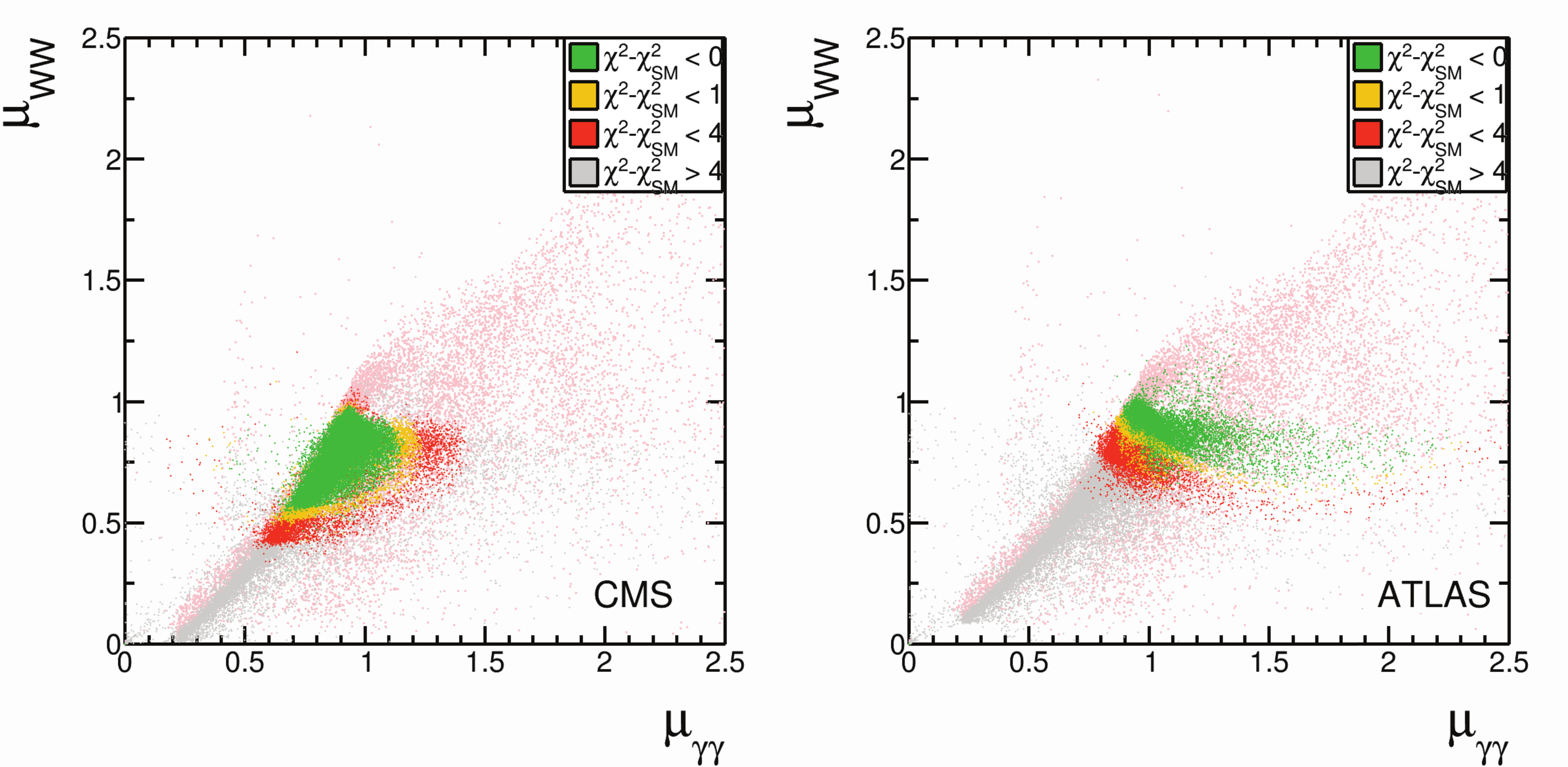}
\end{center}
\caption{General Scenario: Comparison of the $WW^*$ signal strength ratio to the diphoton ratio, using a $\chi^2 - \chi^2_{SM}$ measure to compare the BLH model predictions to the SM predictions, including the full set of measured signal strength ratios.  Parameter points in pink indicate an exclusion at 95\% C.L. due to high mass resonance.}
\label{fig:norm:muyyvsmuww}
\end{figure*}

\begin{figure*}[tp]
\begin{center}
\includegraphics[width=\textwidth]{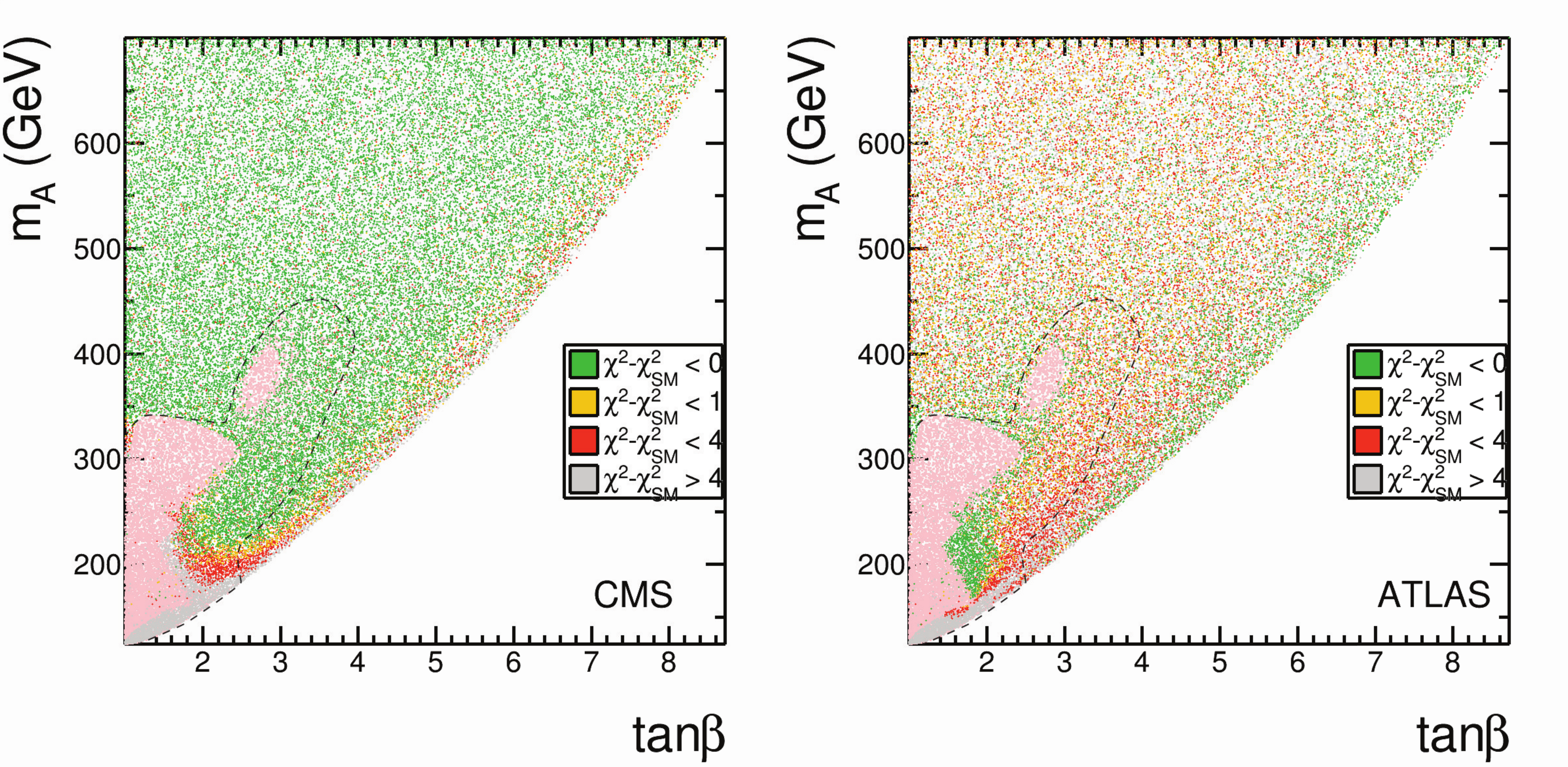}
\end{center}
\caption{General Scenario: Comparison of the CP-odd scalar mass, $m_{A_0}$, and $\tan\beta$ parameters, using a $\chi^2 - \chi^2_{SM}$ measure to compare the BLH model predictions to the SM predictions, including the full set of measured signal strength ratios. Parameter points in pink indicate an exclusion at 95\% C.L. due to high mass resonance. Dashed region indicates approximate boundary enclosing excluded (pink) points.}
\label{fig:norm:tbvsma0}
\end{figure*}

\begin{figure*}[tp]
\begin{center}
\includegraphics[width=\textwidth]{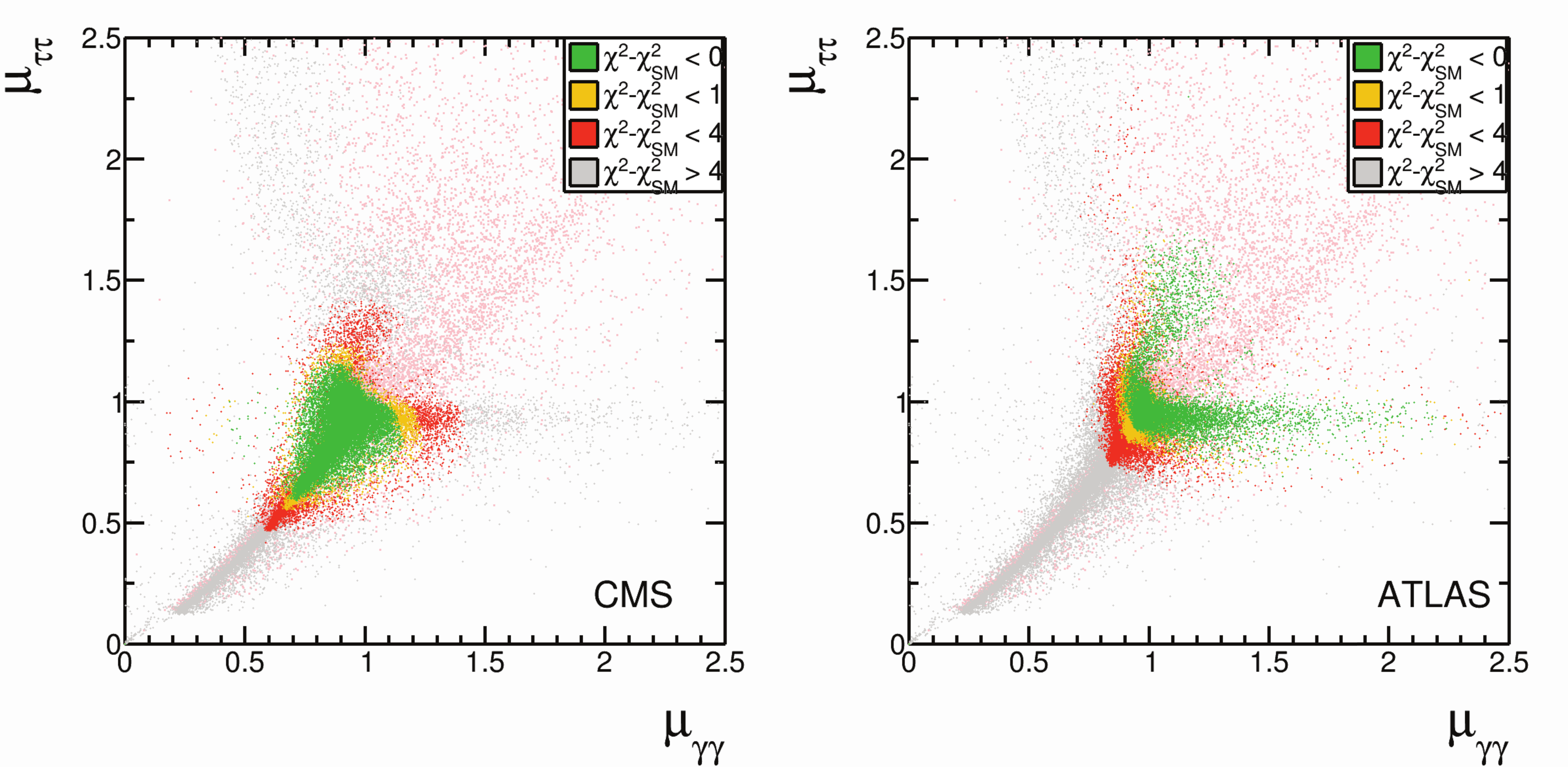}
\end{center}
\caption{General Scenario: Comparison of the $\tau^+\tau^-$ signal strength ratio to the diphoton ratio, using a $\chi^2 - \chi^2_{SM}$ measure to compare the BLH model predictions to the SM predictions, including the full set of measured signal strength ratios. Parameter points in pink indicate an exclusion at 95\% C.L. due to high mass resonance.}
\label{fig:norm:muyyvsmutata}
\end{figure*}

\begin{figure*}[tp]
\begin{center}
\includegraphics[width=\textwidth]{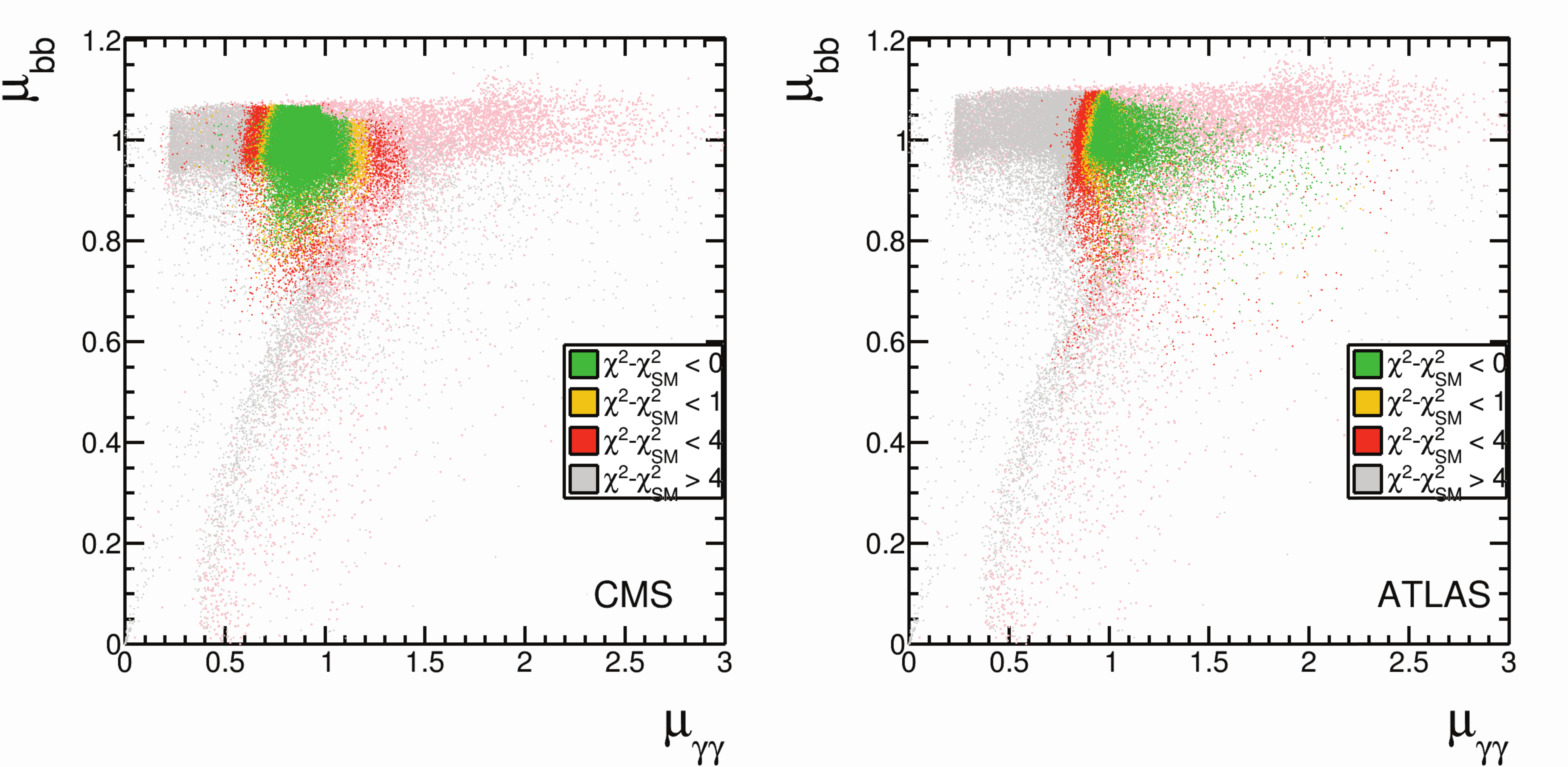}
\end{center}
\caption{General Scenario: Comparison of the $b\bar{b}$ signal strength ratio to the diphoton ratio, using a $\chi^2 - \chi^2_{SM}$ measure to compare the BLH model predictions to the SM predictions, including the full set of measured signal strength ratios. Parameter points in pink indicate an exclusion at 95\% C.L. due to high mass resonance.}
\label{fig:norm:muyyvsmubb}
\end{figure*}

\begin{figure*}[tp]
\begin{center}
\includegraphics[width=\textwidth]{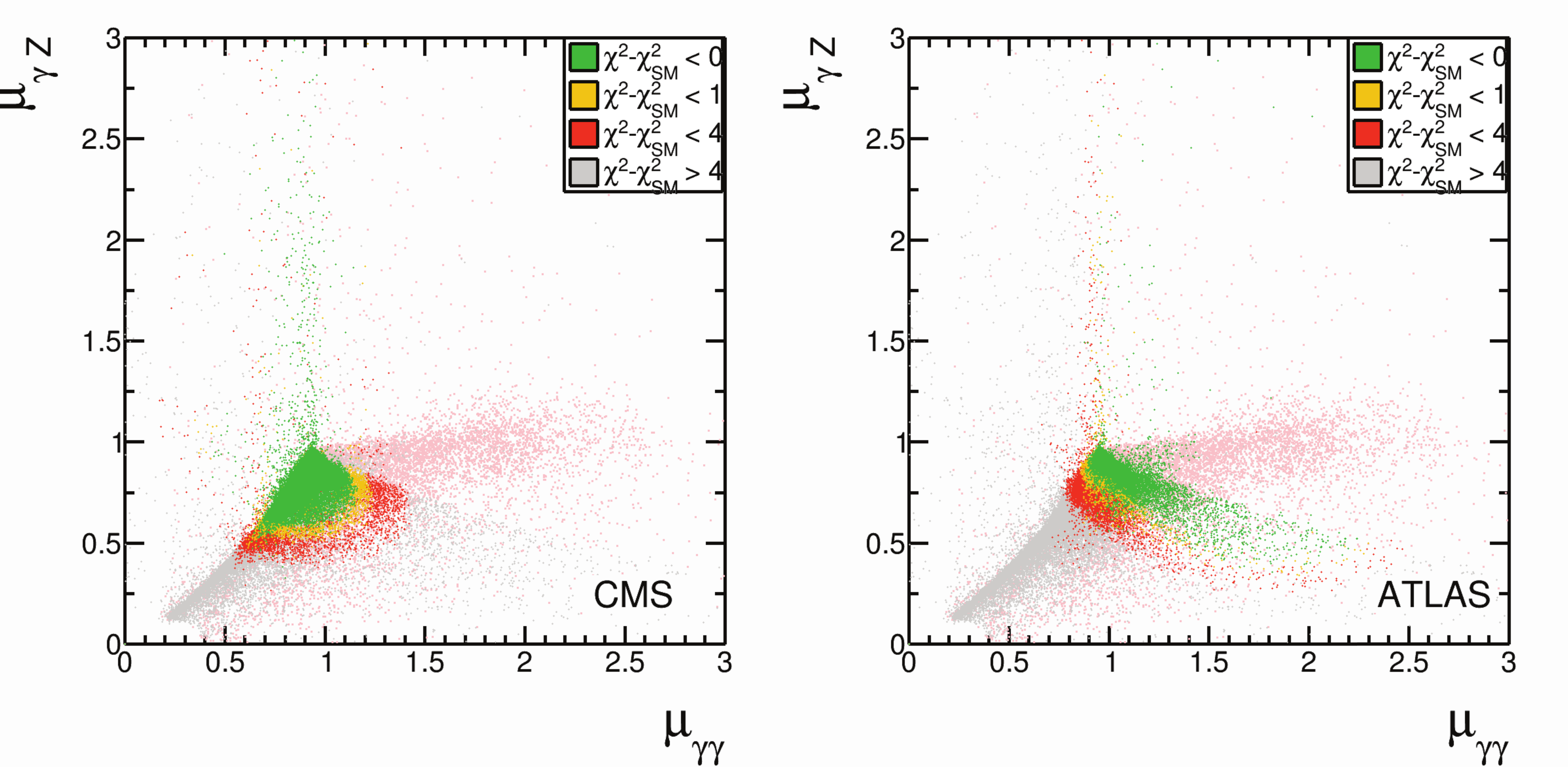}
\end{center}
\caption{General Scenario: Comparison of the $\gamma Z$ signal strength ratio to the diphoton ratio, using a $\chi^2 - \chi^2_{SM}$ measure to compare the BLH model predictions to the SM predictions, including the full set of measured signal strength ratios. Parameter points in pink indicate an exclusion at 95\% C.L. due to high mass resonance.}
\label{fig:norm:muyyvsmuzy}
\end{figure*}

\begin{figure*}[tp]
\begin{center}
\includegraphics[width=\textwidth]{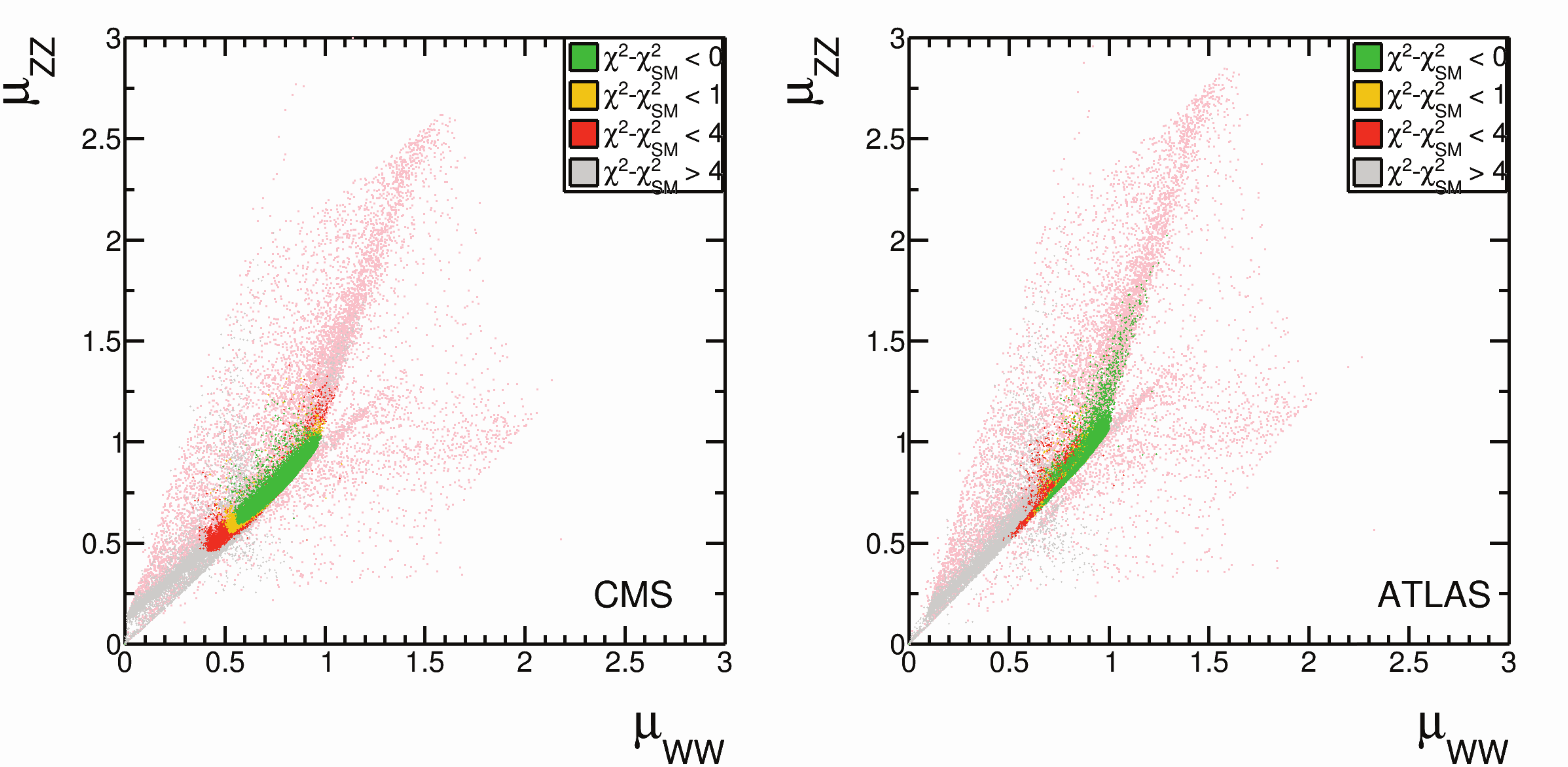}
\end{center}
\caption{General Scenario: Comparison of the $ZZ^*$ and $WW^*$ signal strength ratios, using a $\chi^2 - \chi^2_{SM}$ measure to compare the BLH model predictions to the SM predictions, including the full set of measured signal strength ratios. Parameter points in pink indicate an exclusion at 95\% C.L. due to high mass resonance.}
\label{fig:norm:muwwvsmuzz}
\end{figure*}

In the general scenario, we allow the mass of the CP-odd scalar to vary between the mass of the light Higgs boson and 700 GeV. Thus, there is some overlap between the regions of parameter space explored in this and the near-degenerate scenario. However, since the $\tau^+\tau^-$ results exclude the entirety of the near-degenerate scenario, the points of overlap will be greyed out in Figures \ref{fig:norm:tBvssapb} through \ref{fig:norm:muwwvsmuzz}, which are coloured based on $\chi^2-\chi^2_{SM}$. This is discussed in further detail below.

Figures \ref{fig:norm:tBvssapb} through \ref{fig:norm:muwwvsmuzz} distinctly show that the BLH model is a better fit to the CMS data than is the SM for a significant portion of the parameter space, and is a better fit to the ATLAS data than is the SM for a smaller set of parameter points. The region of better fit occurs for values of $\sin(\alpha+\beta) \gtrsim 0.9$, $m_{A_0} \gtrsim 200$~GeV and $f \gtrsim 1200$~GeV for CMS, as shown on the left side of 
Figures \ref{fig:norm:tBvssapb} and \ref{fig:norm:flvsma0}, respectively. The right side of Figure \ref{fig:norm:flvsma0} indicates that agreement between the BLH model and ATLAS results occurs predominantly for $m_{A_0} \gtrsim 200$~GeV and for $f \gtrsim 2200$~GeV, with an extended region of agreement between $1200 \lesssim f \lesssim 2200$~GeV for parameter sets on the lower boundary of $m_{A_0} \sim 200$~GeV. The favoured large values of $\sin(\alpha+\beta)$ for both experimental result sets are understandable in order to achieve $\mu_{WW} \sim 1$, and larger values of $f$ reduce the contribution from higher order terms in the expansion of $v/f$ in the couplings.

Both the ATLAS and CMS results can be realized in the general scenario, as is evident in Figure \ref{fig:norm:muyyvsmuww}, where we show $\mu_{WW}$ versus $\mu_{\gamma \gamma}$. Figure \ref{fig:norm:tbvsma0} shows that the parameter sets which produce a value of $\mu_{\gamma\gamma} > 1$, as in the ATLAS side of Figure \ref{fig:norm:muyyvsmuww}, occur more frequently near the boundaries of excluded regions of $\tan\beta$ and $m_{A_0}$. In both the CMS and ATLAS cases, the region with $m_{A_0} \lesssim 200$~GeV is mostly ruled out. This means the $A_0$ contribution to $\mu_{\tau\tau}$ is small for parameter points that are allowed and makes that signal strength more consistent with the SM, as shown in Figure \ref{fig:norm:muyyvsmutata}. These results are therefore effectively orthogonal to the near-degenerate scenario, even though there is a small amount of overlap between the parameter spaces generated.

To enhance the excluded (pink) points in Figure \ref{fig:norm:tbvsma0}, a dashed line of the approximate region enclosing the excluded points has been included. For low values of $m_A$ (less than 250 GeV), parameter sets are excluded primarily due to the constraint on $A(H) \rightarrow \gamma\gamma$ production. For larger values of $m_A$, the exclusion comes entirely from the $H_0 \rightarrow W^+W^-$ search. In the areas with overlap between excluded and non-excluded points, there is large variation in the $BR(H_0 \rightarrow W^+W^-)$ and $\sigma(gg \rightarrow H_0)$ values due to influences from the other fundamental parameters (such as $f$ and the scalar sector parameters), and large variations in the total width of the $H_0$. This allows for many parameter sets to avoid exclusion, with sufficiently low  branching ratio or production rate, or both. The small region of pure exclusion at approximately $\tan\beta \sim 3$ and $m_{A_0} \sim 380$~GeV occurs due to all parameter points having a sufficiently large production rate and branching ratio for exclusion. This results in the appearance of an apparently isolated region with a high density of excluded parameter sets. However, this appearance is simply a result of the two dimensional display of values that depend on multiple degrees of freedom in which the excluded points are displayed as the lowest layer.

As is clear from Figures \ref{fig:norm:muyyvsmuww} ($\mu_{\gamma\gamma}$, $\mu_{WW}$), \ref{fig:norm:muyyvsmutata} ($\mu_{\gamma\gamma}$,  $\mu_{\tau\tau}$), \ref{fig:norm:muyyvsmubb} ($\mu_{\gamma\gamma}$, $\mu_{bb}$), and \ref{fig:norm:muyyvsmuzy} ($\mu_{\gamma\gamma}$, $\mu_{\gamma Z}$), the general scenario of the BLH model includes parameter sets that can produce signal strength ratios in better agreement with either CMS or ATLAS than is the SM. The BLH model also allows for the possibility of $\mu_{ZZ} > \mu_{WW}$, as supported by the ATLAS measurement set and visible in the right side of Figure \ref{fig:norm:muwwvsmuzz}. This occurs due to our calculation of an inclusive cross section that includes contributions from the $H_0$, which has larger suppressions to the $W^+W^-$ branching ratio at larger masses than the $ZZ$. An invariant mass windowing of $4l$ events would likely exclude any excess $H_0 \rightarrow ZZ^* \rightarrow 4l$ contributions, and result in a measurement of $\mu_{WW} \sim \mu_{ZZ}$.

\begin{figure*}[tp]
\begin{center}
\includegraphics[width=\textwidth]{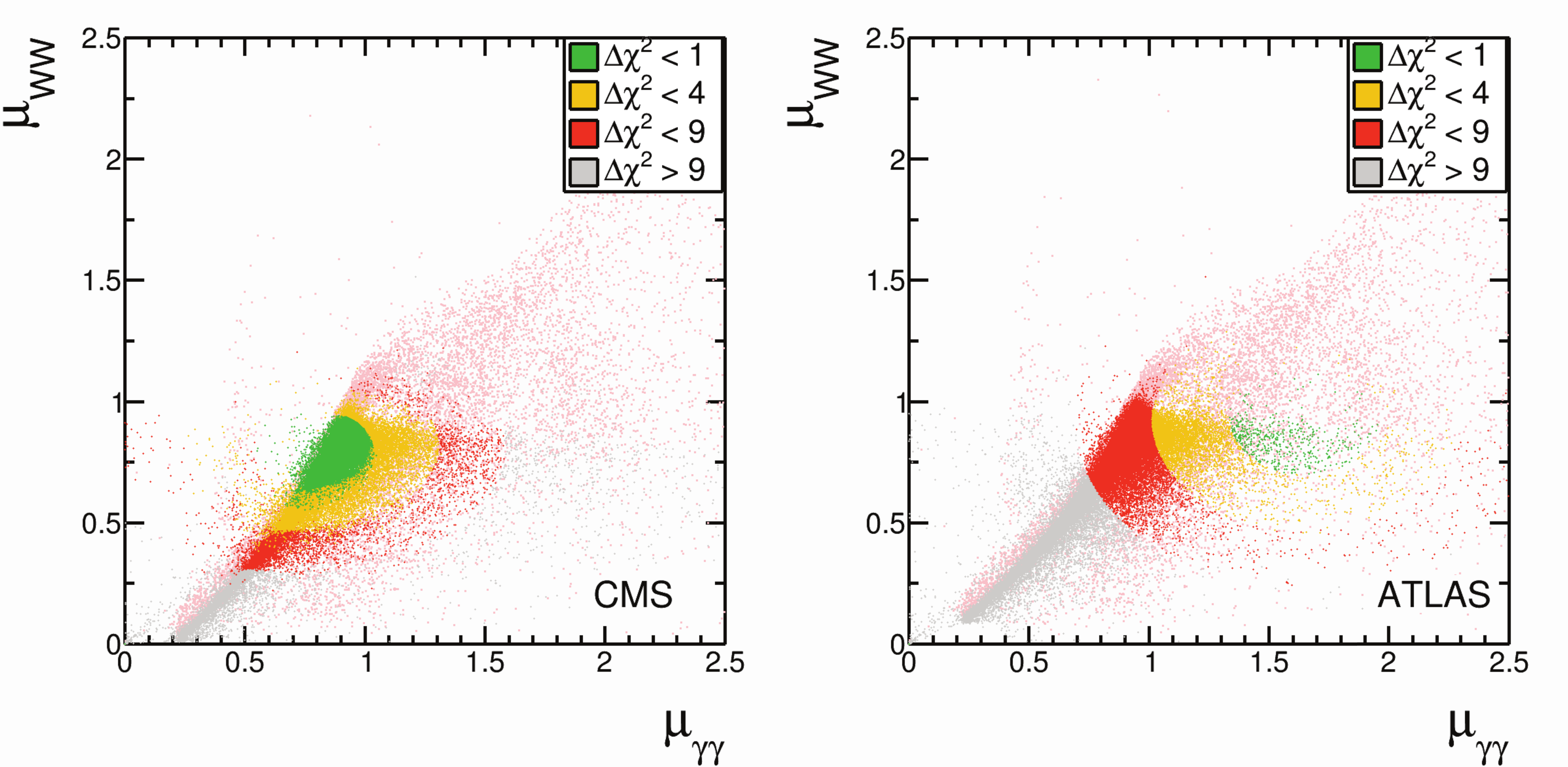}
\end{center}
\caption{General Scenario: Comparison of the $WW^*$ and diphoton signal strength ratios, assuming a reduced $\Delta\chi^2$ calculation (including only $\mu_{\gamma\gamma}$ and $\mu_{WW}$ measurements) to focus on the disparity between the CMS and ATLAS diphoton signal strength ratios. Parameter points in pink indicate an exclusion at 95\% C.L. due to high mass resonance.}
\label{fig:rnorm:muyyvsmuww}
\end{figure*}

It remains to understand the physics underlying the enhancement of the diphoton rate, in agreement with the ATLAS results, in the general scenario. It is not a result of significant contributions from the production and decay of the other neutral Higgs bosons in the 2HDM. While Figures \ref{fig:norm:tBvssapb} through \ref{fig:norm:muwwvsmuzz} show the parameter sets that are in better agreement with the experimental results than is the SM, they do not focus on the parameter sets which best agree with the experimental values. To examine the physics underlying the enhancement in the ATLAS diphoton rate, we consider the reduced $\chi^2$ ($\Delta\chi^2$) that includes only the $\mu_{WW}$ and $\mu_{\gamma\gamma}$ values. This will focus on the results which agree with the diphoton rate while constraining the results to also agree with the precisely measured $WW$ rate. In Figure \ref{fig:rnorm:muyyvsmuww}, we plot $\mu_{WW}$ versus $\mu_{\gamma \gamma}$. Figure \ref{fig:rnorm:muyyvsmuww} shows that the points in agreement using the reduced $\chi^2$ (green points with $\Delta\chi^2 < 1$) are a subset of those that are a better fit to the data than is the SM, as previously shown in Figure \ref{fig:norm:muyyvsmuww} (green points with $\chi^2 - \chi^2_{SM} < 0$).

\begin{figure*}[tp]
\begin{center}
\includegraphics[width=\textwidth]{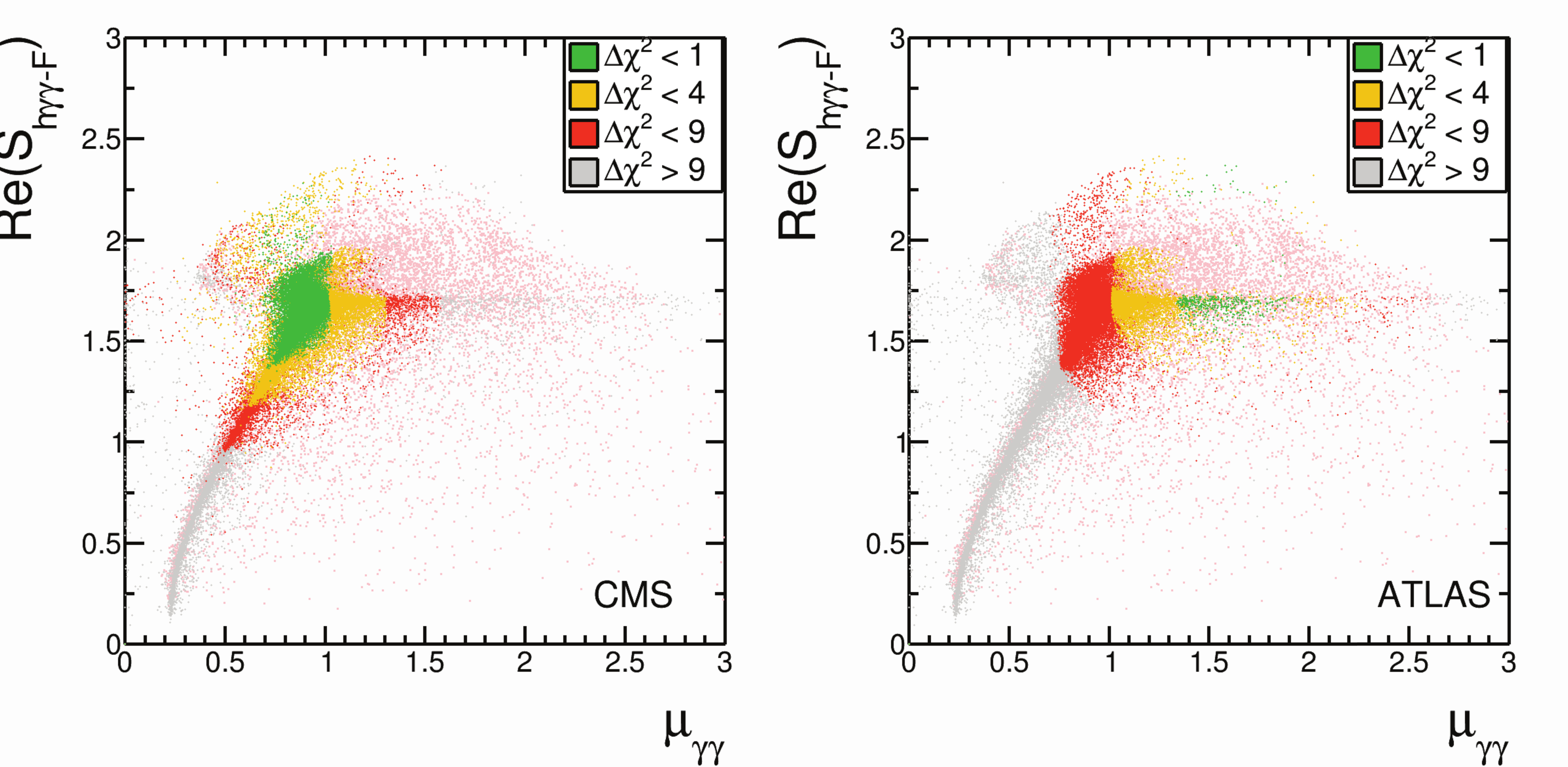}
\end{center}
\caption{General Scenario: Comparison of the fermion contribution to the diphoton effective coupling and the diphoton signal strength ratio, assuming a reduced $\Delta\chi^2$ calculation (including only $\mu_{\gamma\gamma}$ and $\mu_{WW}$ measurements) to focus on the disparity between the CMS and ATLAS diphoton signal strength ratios. Parameter points in pink indicate an exclusion at 95\% C.L. due to high mass resonance.}
\label{fig:rnorm:muyyvsReShyyF}
\end{figure*}

\begin{figure*}[tp]
\begin{center}
\includegraphics[width=\textwidth]{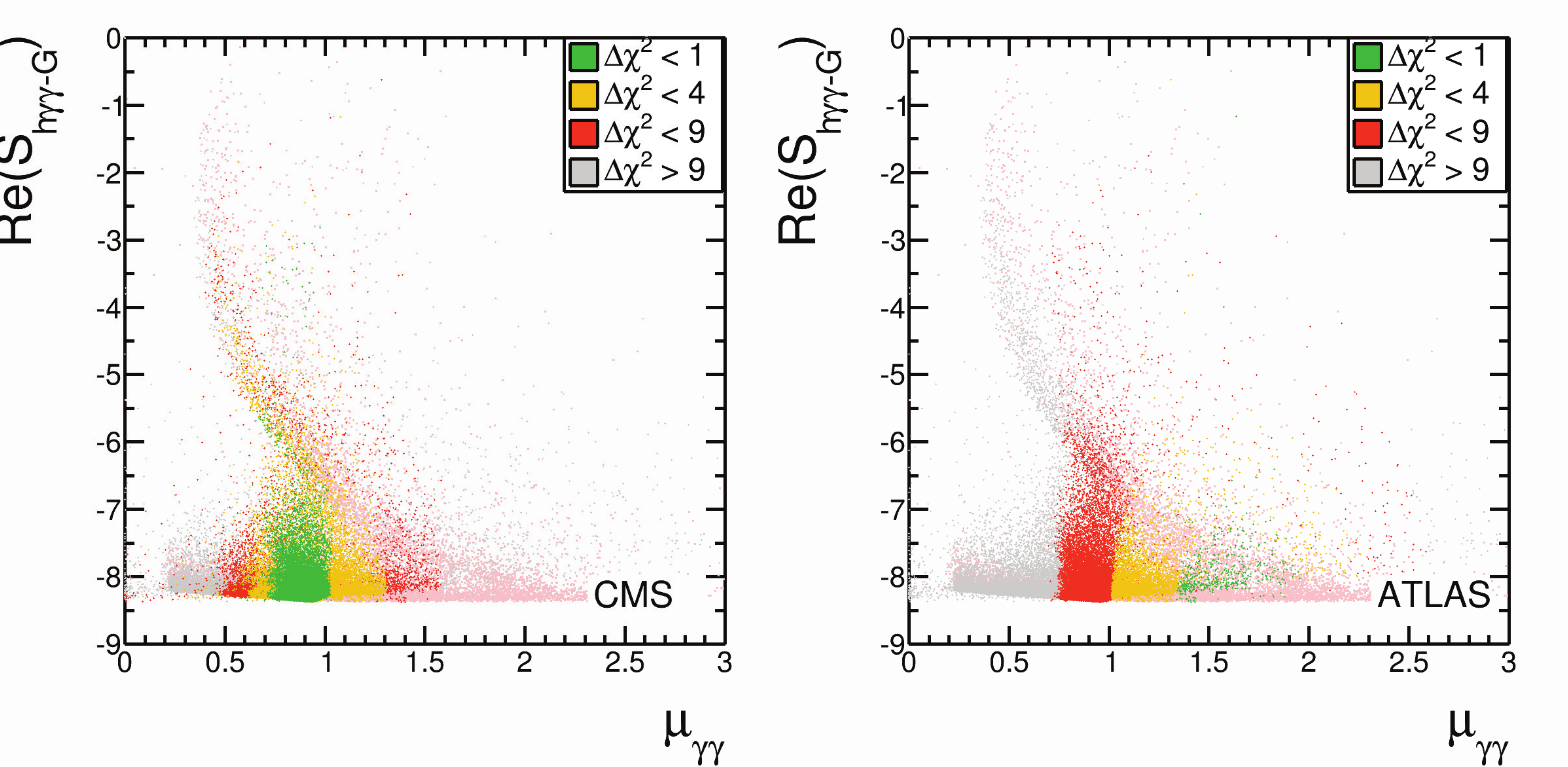}
\end{center}
\caption{General Scenario: Comparison of the gauge boson contribution to the diphoton effective coupling and the diphoton signal strength ratio, assuming a reduced $\Delta\chi^2$ calculation (including only $\mu_{\gamma\gamma}$ and $\mu_{WW}$ measurements) to focus on the disparity between the CMS and ATLAS diphoton signal strength ratios. Parameter points in pink indicate an exclusion at 95\% C.L. due to high mass resonance.}
\label{fig:rnorm:muyyvsReShyyG}
\end{figure*}

\begin{figure*}[tp]
\begin{center}
\includegraphics[width=\textwidth]{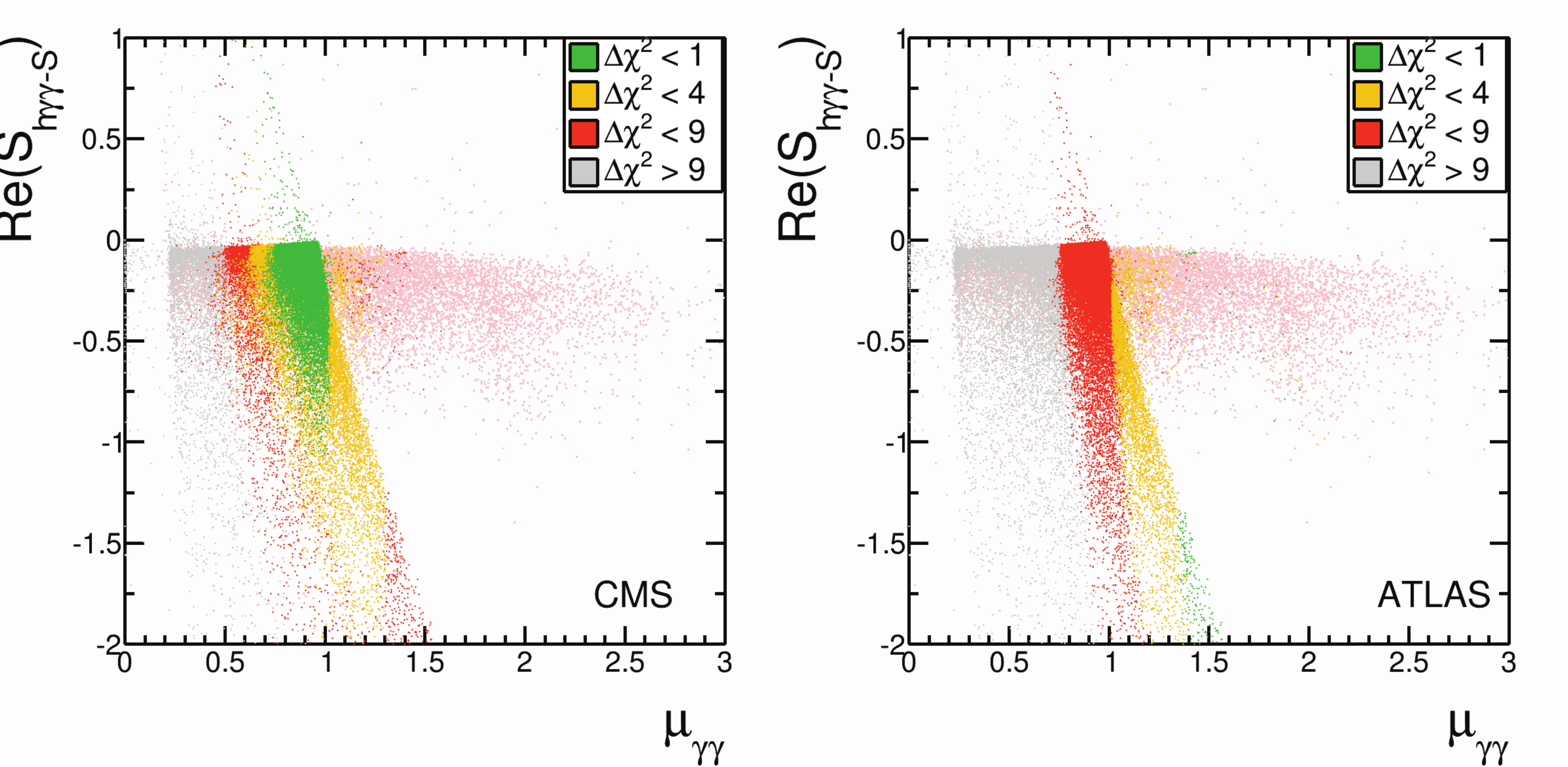}
\end{center}
\caption{General Scenario: Comparison of the scalar contribution to the diphoton effective coupling and the diphoton signal strength ratio, assuming a reduced $\Delta\chi^2$ calculation (including only $\mu_{\gamma\gamma}$ and $\mu_{WW}$ measurements) to focus on the disparity between the CMS and ATLAS diphoton signal strength ratios. Parameter points in pink indicate an exclusion at 95\% C.L. due to high mass resonance.}
\label{fig:rnorm:muyyvsReShyyS}
\end{figure*}

We now investigate the contributions of different particles to the diphoton loop process. By separately examining the real component of the fermion, scalar and gauge contribution to the diphoton effective coupling, $S_{h\gamma\gamma-F}=\sum_f A_{f,\gamma\gamma}^{BLH}$, $S_{h\gamma\gamma-G}=\sum_V A_{V,\gamma\gamma}^{BLH}$,  and $S_{h\gamma\gamma-S}=\sum_S A_{S,\gamma\gamma}^{BLH}$  respectively, as in Figures \ref{fig:rnorm:muyyvsReShyyF}, \ref{fig:rnorm:muyyvsReShyyG} and \ref{fig:rnorm:muyyvsReShyyS}, it becomes clear that there are two possibilities that result in a significant enhancement of the diphoton rate. As shown in Figure \ref{fig:rnorm:muyyvsReShyyF} (right panel), the contribution from fermions is approximately SM-like. This is as expected since significant deviation would also result in significant alterations of the gluon fusion effective coupling to the Higgs. Similarly, as shown in Figure \ref{fig:rnorm:muyyvsReShyyG} (right panel), the contribution from gauge bosons is also SM-like.  The additional heavy gauge bosons of the BLH model do not affect the results due to their large mass and suppressed couplings. Any significant alteration of the $hWW$ coupling from its SM value would result in disagreement with the $\mu_{WW}$ results.

\begin{figure*}[tp]
\begin{center}
\includegraphics[width=\textwidth]{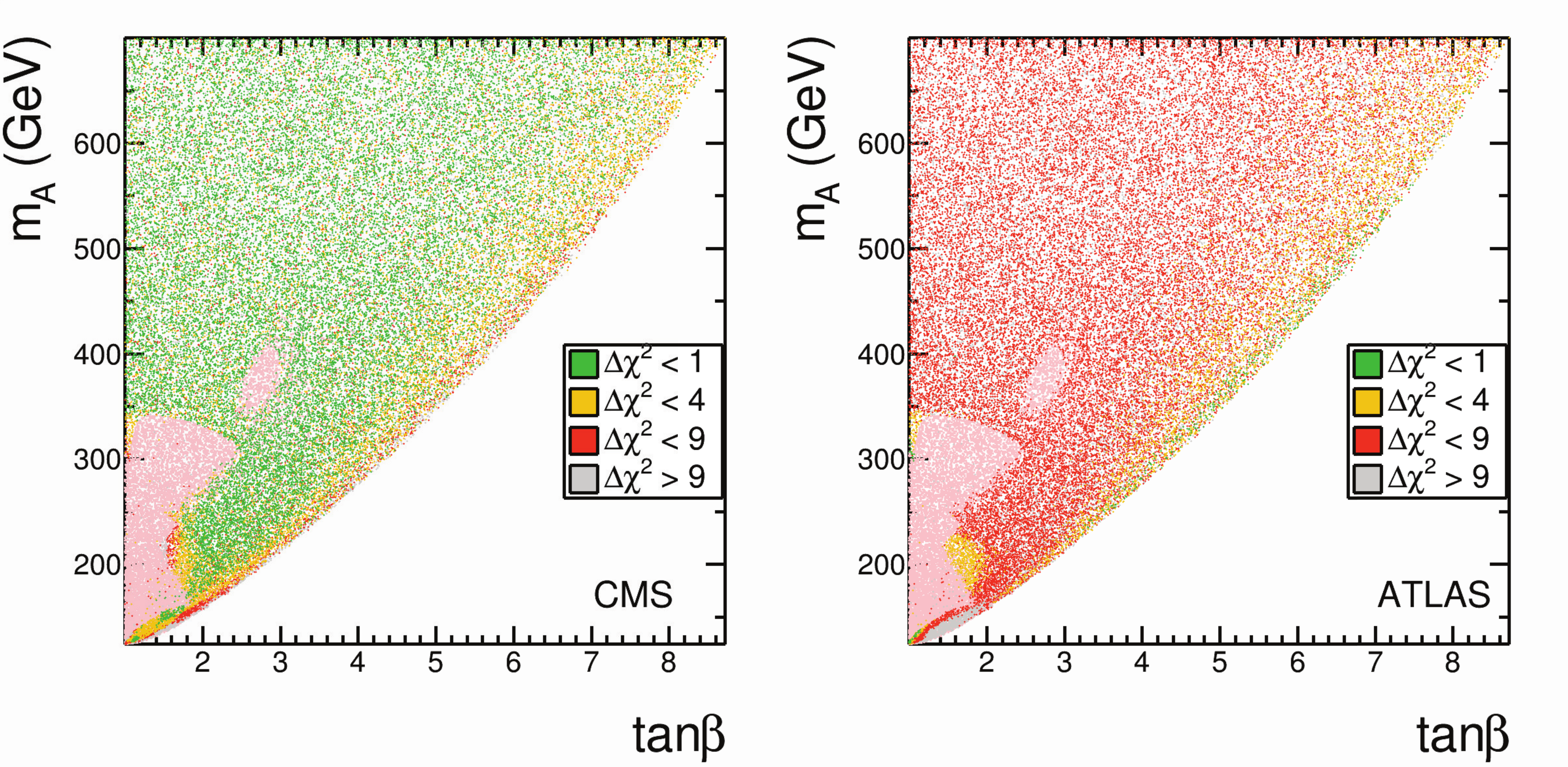}
\end{center}
\caption{General Scenario: Comparison of the CP-odd scalar mass, $m_{A_0}$, and $\tan\beta$ parameters, assuming a reduced $\Delta\chi^2$ calculation (including only $\mu_{\gamma\gamma}$ and $\mu_{WW}$ measurements) to focus on the disparity between the CMS and ATLAS diphoton signal strength ratios. Parameter points in pink indicate an exclusion at 95\% C.L. due to high mass resonance.}
\label{fig:rnorm:tBvsma0}
\end{figure*}

The right panel of Figure \ref{fig:rnorm:muyyvsReShyyS} shows the most obvious source of enhancement of the diphoton rate, as measured by ATLAS. The enhancement occurs when the scalar contribution to the diphoton effective coupling becomes significant, dominated by the contribution from the charged Higgs field, $H^\pm$. The dominant component of the diphoton effective coupling comes from the $W$ boson loop, which has a negative relative value, as do the contributions from the charged scalars ($H^\pm$, $\eta^\pm$ and $\phi^\pm$), while the sum of all of the fermion loops contribute positively. Thus, enhancement of the scalar loop contribution increases the effective coupling strength of $h\gamma\gamma$ (reductions of the contribution from fermion loops would have a similar effect, but are not important here). 
We have determined that the diphoton enhancement occurs where the $h_0 H^+ H^-$ coupling becomes large for
large values of $\lambda_0$, corresponding to where $\tan\beta$ is near the the upper boundary of its allowed range
as seen in the right panel of Figure \ref{fig:rnorm:tBvsma0}.
This region lies close to the border of perturbativity constraints, as discussed in Eq. \ref{eq:tBmax}.

\begin{figure*}[tp]
\begin{center}
\includegraphics[width=\textwidth]{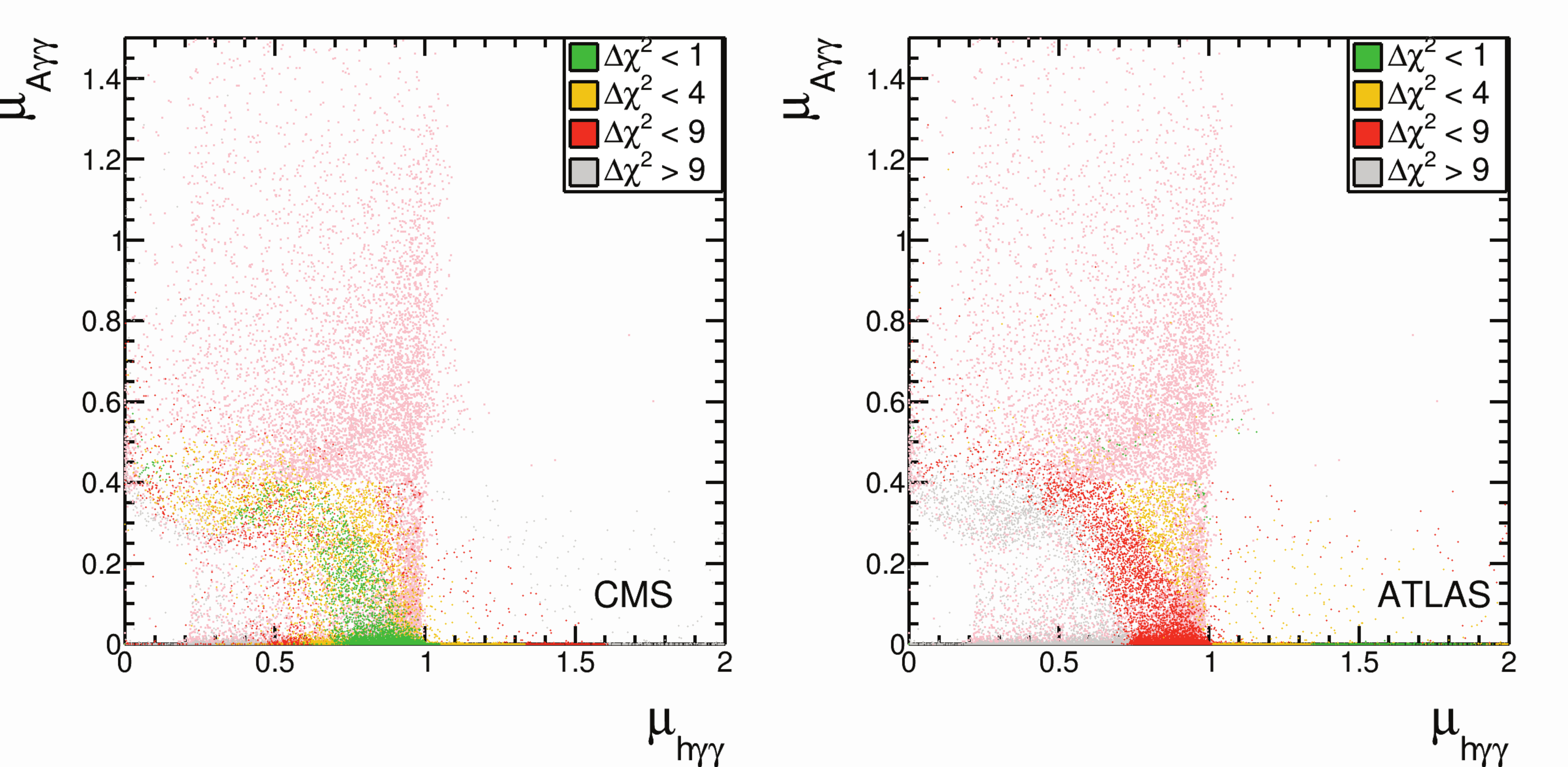}
\end{center}
\caption{General Scenario: Comparison of the pseudoscalar contribution to the diphoton signal strength ratio to the light Higgs boson contribution, assuming a reduced $\Delta\chi^2$ calculation (including only $\mu_{\gamma\gamma}$ and $\mu_{WW}$ measurements) to focus on the disparity between the CMS and ATLAS diphoton signal strength ratios. Parameter points in pink indicate an exclusion at 95\% C.L. due to high mass resonance.}
\label{fig:rnorm:hmuyy1vsAmuyy1}
\end{figure*}

Figure \ref{fig:rnorm:hmuyy1vsAmuyy1} also shows several parameter points which fit the data well but do not have a significant scalar loop contribution to the diphoton effective coupling. These points correspond to a moderate contribution to the diphoton rate from the production of the CP-odd scalar at masses above 300 GeV. This is an artifact of the method we use for calculating, which determines an inclusive ($\sigma_{h_0}+\sigma_{H_0}+\sigma_{A_0}$) cross section, but excludes parameter sets only for which the heavier resonances would be distinguishable at 95\% C.L. In other words, if the heavier resonances do not result in an exclusion, they contribute to the total cross section calculated. Little information was given by the experiments regarding any invariant mass windowing incorporated into the determination of the diphoton excess, and so we chose an inclusive cross section calculation.

\begin{figure*}[tp]
\begin{center}
\includegraphics[width=\textwidth]{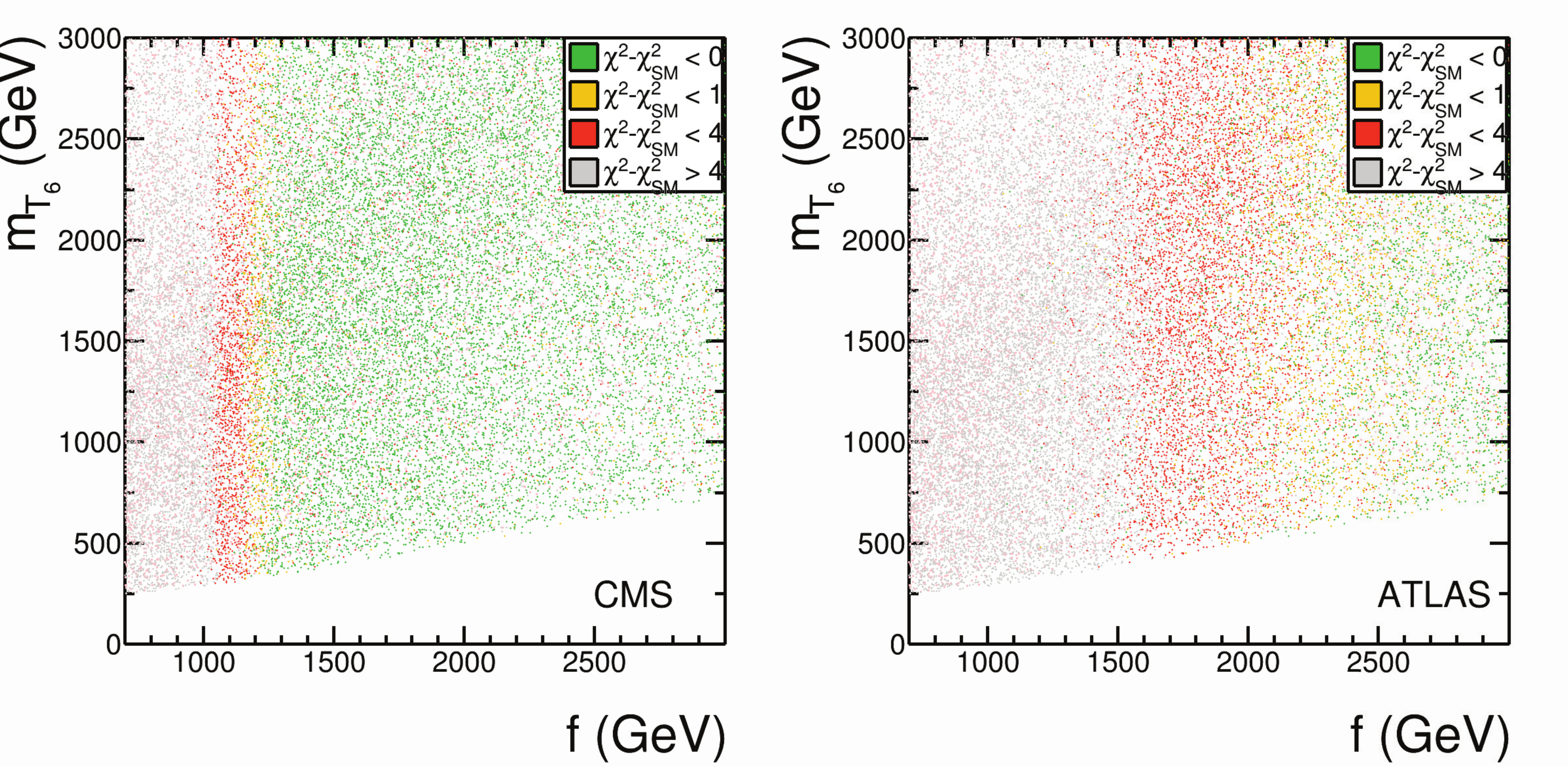}
\end{center}
\caption{General Scenario: Comparison of the mass of the $T_6$ heavy vector-like quark to the $f$ parameter, using a $\chi^2 - \chi^2_{SM}$ measure to compare the BLH model predictions to the SM predictions, including the full set of measured signal strength ratios. Parameter points in pink indicate an exclusion at 95\% C.L. due to high mass resonance.}
\label{fig:norm:flvsmT6}
\end{figure*}

With regards to the other parameters in the model, particularly the heavy quark mixing angles, $\theta_{12}$ and $\theta_{13}$, and heavy gauge boson mixing angle, $\theta_g$, no constraints can be placed with the existing data. This is because these states do not get a large component of their mass from the Higgs vacuum expectation value, and so they do not contribute significantly to the loop factors. In Figure \ref{fig:norm:flvsmT6}, we show, as an example, the $T_6$ mass versus $f$. The only constraint that can be determined is an overall mass constraint arising directly from the constraint on $f$. In particular, the minimum heavy quark mass as determined from the Higgs data is approximately 300 GeV - the precise value of which is unimportant, as direct constraints from pair production searches for heavy vector-like quarks rule out much heavier states \cite{Chatrchyan:2013uxa,TheATLAScollaboration:2013sha}.

\section{Summary \label{sec:summary}}

The BLH model 
provides, in principle, a rich source of phenomenology. Apart from incorporating a two Higgs doublet model, it includes additional scalar triplets, heavy gauge bosons, and a set of six new heavy quarks. In this paper, we have investigated whether the model includes parameter sets that are consistent with the Higgs boson signal strength ratios recently measured by the CMS and ATLAS experiments. We have used a couple of $\chi^2$ measures, one to compare the fit of the BLH model to the data relative to that of a SM fit and another to identify regions of BLH model parameter space favoured by the results of the two experiments. As described above, we have found that the BLH model can reproduce the results of either experiment but primarily via modifications to the couplings of top quarks to the Higgs states and through the contributions of additional scalar states in the context of a 2HDM. As a 2HDM, the BLH model provides two possible scenarios: the general case in which a single Higgs state dominates contributions to the signal strength ratios, and the near-degenerate case in which multiple Higgs states contribute to the observed results.

At this time, the experimental data remains statistically limited and there remain discrepancies between the central values of the CMS and ATLAS results. However, using the full set of measured signal strength ratios, we find that the BLH model in the near-degenerate scenario is a worse fit to the data of each experiment than is the SM for almost all parameter sets. This is a consequence of a large enhancement of the $\mu_{\tau\tau}$ signal strength by the contribution of the CP-odd scalar state nearly degenerate with the light Higgs boson. In this scenario, for parameter sets consistent with the observed diphoton rates, $\mu_{\tau\tau}$ is predicted to be three to five times larger than its SM value, and several sigma larger than the value measured by CMS or ATLAS. Consequently, a precise measurement of $\mu_{\tau\tau}$ will be sufficient to exclude the near-degenerate scenario in the BLH model.

On the other hand, large regions of the general BLH parameter space provide a better fit to the experimental results than does the SM. This corresponds to $\sin(\alpha+\beta) \gtrsim 0.9$ in order to achieve $\mu_{WW} \sim 1$, while $f \gtrsim 1200$ GeV is necessary such that higher order corrections in the expansion in $v/f$ do not reduce couplings between the $t$ quark and $h_0$, and the $W$ boson and $h_0$. A CP-odd scalar mass of $m_{A} \gtrsim 300$ GeV is favoured, resulting in a value of $\mu_{\tau\tau}$ that is consistent with the SM. These rather general constraints provide good agreement with the CMS diphoton results, allowing a large range of $\tan\beta$ and $m_A$ values. The signal strength ratios are produced primarily through the light Higgs boson with approximately SM-like couplings.

Due to the enhancement of the overall scalar and CP-odd scalar production for $130 \lesssim m_A \lesssim 300$~GeV, much of this region of parameter space is already directly ruled out at 95\% C.L. by heavy Higgs searches in the $WW$ and $\gamma\gamma$ channels. Additionally, all parameter regions of the BLH model predict the values of $\mu_{WW}$ and $\mu_{ZZ}$ to be similar, such that the entirety of the BLH model would be excluded if the difference in the $\mu_{ZZ}$ and $\mu_{WW}$ results becomes statistically significant to the degree currently measured by ATLAS.

It is possible to reproduce the ATLAS $\mu_{\gamma\gamma}$ measurement with a SM-like $\mu_{\tau\tau}$ in the BLH model, but for a restricted space of parameter sets. The physical origin of the enhancement of the diphoton rate is a significant enhancement of the charged Higgs field ($H^\pm$) contribution to the diphoton loop. The diphoton enhancement occurs for maximal values of $\tan\beta$, where the $h_0 H^+ H^-$ coupling becomes large.

More accurate measurements, including the $b\bar{b}$, $\tau^+\tau^-$ and $Z\gamma$ final states, with higher luminosity will be crucial to determining the status of the BLH model, and for determining the values of the scalar sector parameters. The 2HDM sector of the BLH model is likely its most accessible aspect, with fairly light CP-odd and, consequently, charged Higgs states allowed. Discovery and measurement of heavy quark partners can lead to further constraints on the value of the scale $f$, as discussed in \cite{Godfrey:2012tf}. However, in the BLH model, measurements of the mass and branching ratios of the lightest heavy quark partner is insufficient to significantly constrain the value of $f$, due to the presence of degenerate states and the involvement of the two mixing angles, $\tan\theta_{12}$ and $\tan\theta_{13}$. In \cite{Martin:2012}, it was shown that measurement of several of the heavy quark masses is needed to isolate the value of the scale $f$ from the values of the heavy quark mixing angles.

\acknowledgments

The authors would like to thank Heather Logan, Thomas Gr\'egoire, Stephen Godfrey, Michael Spira, and Peter Winslow for guidance and assistance in this project. The authors would also like to apologize to M.~Schmaltz, D.~Stolarski and J.~Thaler for not using the full name of their model from \cite{Schmaltz:2010ac} in our work. This research was supported in part by the Natural Sciences and Engineering Research Council of Canada.  
  

\end{document}